\documentclass[fleqn,usenatbib,useAMS]{mnras}
\usepackage{amsmath,amssymb}
\usepackage{xspace}
\usepackage{graphicx}
\usepackage{orcidlink}
\usepackage{comment}

\defcitealias{Schmidt20}{Schmidt \& Malz, et al.}

\newcommand{\citeDCt}{\citetalias{Schmidt20} (\citeyear{Schmidt20})\xspace}

\newcommand{\refresponse}[1]{{\color{black}{#1}}}
\newcommand{\revone}[2]{{\color{black}{#2}}}
\newcommand{\refresponsetwo}[1]{{\color{black}{#1}}}
\newcommand{\cwr}[1]{{\color{black}{#1}}}

\title[Diagnosing Spec-$z$ Imperfection on Photo-$z$]{Diagnosing the Effects of \cwr{Spectroscopic Training Set Imperfection} on Photometric Redshift Performance}

\author[]{%
Alice~Crafford\orcidlink{0009-0006-4576-2588},$^{1}$
Alex~I.~Malz\orcidlink{0000-0002-8676-1622},$^{2,1}$
Tianqing~Zhang\orcidlink{0000-0002-5596-198X},$^{1,3}$
Rachel~Mandelbaum\orcidlink{0000-0003-2271-1527},$^{1}$\thanks{E-mail: rmandelb@andrew.cmu.edu}
Olivia~Lynn\orcidlink{0000-0001-5028-146X},$^{1}$ \newauthor
Federico~Berlfein\orcidlink{0009-0004-4113-9938},$^{1}$
Johann~Cohen-Tanugi\orcidlink{0000-0001-9022-4232},$^{4}$
John~Franklin~Crenshaw\orcidlink{0000-0002-2495-3514},$^{5,6,7}$
Qianjun~Hang,$^{8}$\newauthor
Irene~Moskowitz\orcidlink{0000-0002-2206-8589},$^{9}$
Drew~Oldag\orcidlink{0000-0001-6984-8411},$^{10}$
Samuel~J.~Schmidt\orcidlink{0000-0002-5091-0470},$^{11}$
Ziang~Yan,$^{12,13}$\newauthor
and the LSST Dark Energy Science Collaboration
\\
$^{1}$McWilliams Center for Cosmology and Astrophysics, Department of Physics, Carnegie Mellon University, Pittsburgh, PA, USA\\
$^{2}$Space Telescope Science Institute, Baltimore, MD, USA\\
$^{3}$Department of Physics and Astronomy and PITT PACC, University of Pittsburgh, Pittsburgh, PA 15260, USA\\
$^{4}$Laboratoire de Physique de Clermont Auvergne, Universit\'e Clermont Auvergne and CNRS, Clermont-Ferrand, France\\
$^{5}$Kavli Institute for Particle Astrophysics and Cosmology, Stanford University, Stanford, CA, USA\\
$^{6}$Department of Physics, Stanford University, Stanford, CA 94305, USA\\
$^{7}$SLAC National Accelerator Laboratory, 2575 Sand Hill Road, Menlo Park, CA 94025, USA\\
$^{8}$Department of Physics \& Astronomy, University College London, Gower Street, London WC1E 6BT, UK\\
$^{9}$Department of Physics and Astronomy, Rutgers, The State University of New Jersey, Piscataway, NJ 08854, USA\\
$^{10}$DIRAC Institute, Department of Astronomy, University of Washington, Seattle, WA 98195, USA\\
$^{11}$Department of Physics, University of California, One Shields Avenue, Davis, CA 95616, USA\\
$^{12}$Faculty of Physics and Astronomy, Astronomical Institute (AIRUB), Ruhr University Bochum, German Centre for Cosmological\\Lensing, 44780 Bochum, Germany\\
$^{13}$Graduate School of Science, Nagoya University, Furocho, Chikusa-ku, Nagoya, Aichi 464-8602, Japan
}

\begin{document}

\maketitle

\begin{abstract}%
Most LSST extragalactic science will rely on photometric redshifts (photo-$z$) to extract distance information for the galaxies. 
However, an incomplete or non-representative training set can introduce bias into photo-$z$ estimation. 
It is necessary to understand how various forms of training set imperfection, such as incompleteness and non-trivial spectroscopic target selection, affect photo-$z$ estimation algorithms, and to identify metrics best-suited to quantify the impact.
This work aims to systematically study metrics for diagnosing how various photo-$z$ methods react to certain types of training set incompleteness and non-representativeness. 
We use methods available through the open-source Python library Redshift Assessment Infrastructure Layers (\texttt{RAIL}; \citealt{2025arXiv250502928T}) to systematically test the algorithms \texttt{CMNN}, \texttt{GPz}, \texttt{FlexZBoost}, and \texttt{PZFlow} on mock training data degraded in accordance with several existing spectroscopic sky surveys, as well as under conditions of inverse redshift incompleteness\refresponse{, which approximately mimics observed patterns of incompleteness at high redshift}. 
We employ the algorithm \texttt{TrainZ} as a control.
Finally, we quantify photo-$z$ algorithm performance using a variety of statistical metrics  implemented externally to RAIL. 
We determine that the Kullback-Liebler Divergence, Wasserstein Distance, and Probability Integral Transform are particularly informative metrics with which to assess the impact of training set imperfection on algorithmic performance. 
We also find that inverse redshift incompleteness effects alone lack the complexity to realistically represent anticipated training data. 
\end{abstract}


\begin{keywords}
  galaxies: distances and redshifts -- methods: statistical
\end{keywords}

\section{Introduction} \label{sec:intro}

The Vera C.~Rubin Observatory Legacy Survey of Space and Time (LSST, \citealt{ivezic2019}) will collect the largest-ever imaging survey dataset in the next decade. LSST will detect and observe $\sim 10^{9}$ galaxies\revone{ in the universe}{}, which is transformative for many extragalactic science cases. 
Spectroscopic measurements will not be available for most galaxies in LSST due to its depth and sheer data volume; 
therefore, it will be necessary to estimate the redshift of the galaxies based on their photometry. 
Such techniques are referred to as photometric redshifts, or photo-$z$ \citep[for a review, see][]{2019NatAs...3..212S,2022ARA&A..60..363N}. 

Most \refresponse{data-driven photo-$z$ algorithms} require a training set (photometry cross-matched with precise redshift measurements from spectroscopic surveys). 
Because photo-$z$ point estimates are subject to nontrivial statistical and systematic uncertainties, redshifts are best characterized by posterior probability density functions (PDFs) that enable quantification of non-Gaussian error;
the designation of posterior conveys that data-driven algorithms typically estimate redshifts conditioned on photometric data rather than the probability of the observed photometry conditioned on redshift, which is only possible with a physical model.
\citeDCt explored the quality of photo-$z$ PDFs for different estimators under idealized conditions, i.e.~a perfectly representative and complete training set of generous size\revone{}{ yet nonetheless uncovered several major areas for improvement in the field of photo-$z$}. 

\revone{However}{Most obviously}, a realistic training set will likely never be a representative subsample as the faint galaxy samples used for cosmological measurements in LSST, due to different sensitivities of spectroscopic instruments along with the spectroscopic survey selection functions \citep{Stylianou22}, \refresponse{and the specific training set conditions to which various estimation algorithms are sensitive or performant is not yet adequately understood. } \cwr{The mismatch between the training set and the photometric sample to which the photometric redshift estimators are to be applied induces a statistical effect known as `covariate shift'.} 
This experiment picks up from \citeDCt by testing empirical estimators under realistically imperfect assumptions of training set size, completeness, and representativity \cwr{to assess their sensitivity to this covariate shift}.

\revone{}{However, the other findings of \citeDCt were more subtle.
First, the most commonly used metrics of photo-$z$ performance were shown to \refresponsetwo{have a failure mode wherein they responded favorably to a pathological approach to photo-$z$ that does not use the galaxy photometry.  Since the key factor} we wish to assess about a given estimator and prior information \refresponsetwo{is how well they extract redshift information from photometry, this finding highlighted} a need for exploration and vetting of potential performance metrics for the purpose of identifying promising estimators and characterizing their sensitivity to training set imperfection.
Second, the statistical metrics available for exploration depend on what truth information is accessible;
if true redshifts are known, they may be compared to photo-$z$ point estimates, but without true posterior PDFs, there are limited options for evaluating estimated posterior PDFs.}

\revone{}{This work is a follow-up to \citeDCt, but it is not a data challenge aiming to vet estimators against one another;
rather, we aim to inform the way in which such a vetting should be done in order to best select and optimize photo-$z$ algorithms.
To that end, our experiments focus on the metrics by which photo-$z$ estimation is assessed, rigorously testing those against one another in controlled experiments of training set imperfection on a non-exhaustive subset of estimators, considering the sensitivity of the metrics to the degree and type of imperfection in relation to the differences between algorithm types.
The metric(s) used to choose and tune algorithms for various science cases should reflect the inherent and thus anticipated worsening of redshift information recovery as data quality drops regardless of the choice of estimator\refresponsetwo{. I}t should be possible to interpret an estimator's favorable metric value in terms of some information-maximizing property of the \refresponsetwo{estimator} itself rather than a \refresponsetwo{failure of the metric to respond well to changes in the estimator's performance due to some quirk in how the metric was designed}.}

Our work uses the open-source Python package Redshift Assessment Infrastructure Layers (RAIL\revone{}{; \citealt{2025arXiv250502928T}})\revone{,  to facilitate}{ that was designed to enable the photo-$z$ community to address the concerns highlighted by \citeDCt by facilitating} data generation and large-volume analysis of photo-$z$ algorithms. 
RAIL provides a library of modules to generate mock photometric catalogs as training sets, a collection of data-driven and template-fitting algorithms to estimate photo-$z$ PDFs, and statistical metrics to evaluate the performance of individual runs. 
All RAIL functionality is written into stages, with the vision that users may connect different stages and turn them into an end-to-end data processing pipeline. 

One of the capabilities of RAIL is to generate a realistic photometric catalog through its ``creation'' modules. 
The creation modules have two main functionalities: (a) the engines emulate the photometry-redshift relation learned from sophisticated simulations, such as \citet{Korytov2019} and \citet{Troxel2023}; (b) the degraders add LSST-like noise and apply spectroscopic selections to the truth catalog to produce a realistic catalog for training and testing purposes. 
These functionalities enable the study of the training set incompleteness and non-representativity effects on the \refresponse{photo-$z$}, as its more extensive collection of tunable degradation stages allows us to approach the realistic complexity of the datasets that will be collected by LSST.  
RAIL \revone{is also}{also provides a shared API to access} a library of more than a dozen photo-$z$ estimation algorithms, including data-driven, template-fitting, and image-based algorithms. 
In this study, we apply five different photo-$z$ estimation algorithms implemented in RAIL on the degraded training set, to study their sensitivity to different types and levels of imperfection in the training set. 

In this study, we are interested in different statistical metrics that help quantify the imperfections in photo-$z$ due to imperfect training sets\revone{.  
We are particularly interested in}{, especially} the metrics that compare the estimated posterior PDF to the true posterior PDF of the galaxy, which is defined as the conditional probability \refresponsetwo{distribution} of the redshift given the photometry \revone{}{in the absence of bias due to an estimator or prior information;
this quantity was inherently unavailable in the study of \citeDCt but is accessible for photometry forward-modeled by RAIL's creation functionality and critical to evaluating estimators of photo-$z$ PDFs}.  
We apply seven such metrics to the estimated PDF to study their behavior\revone{.}{, aiming to determine whether they are suitable for assessing the sensitivity of estimators to imperfect prior information and for comparing estimators to one another.
A good metric should be sensitive to how well an estimator extracts redshift information from imperfect photometry under imperfect prior information, demonstrating sensitivity to when either of these inputs worsens, for any estimator;
if we can draw conclusions demonstrating that it holds for the choice of estimator, our findings can inform not only assessments of training set requirements but also optimization of the values of tuning parameters of an estimator and even the choice of estimator itself.}  
\refresponse{In all cases, we are quantifying the performance of individual galaxy photo-$z$ with basic statistical methods.  
This sets the stage for future studies that must also consider science-driven metrics and the performance for galaxy ensembles rather than individual photo-$z$.}

The paper is organized as follows. Section~\ref{sec:methodsdata} introduces the methodology, including the data generation, degradation, and photo-$z$ estimation methods.  It also describes the data used for training the truth catalog emulator.  Section~\ref{sec:metrics} describes the different statistical methods used to \refresponse{quantify} photo-$z$ quality, and provides some intuition regarding their interpretation.  Section~\ref{sec:results} describes the results of our experiments and our observation of how different patterns in training set degradation are reflected in different statistical metrics for each photo-$z$ algorithm. In Section~\ref{sec:conclusion}, we discuss our conclusions from these experiments, and their implications for the future.



\begin{figure*}
  \centering
  \includegraphics[width=0.75\textwidth]{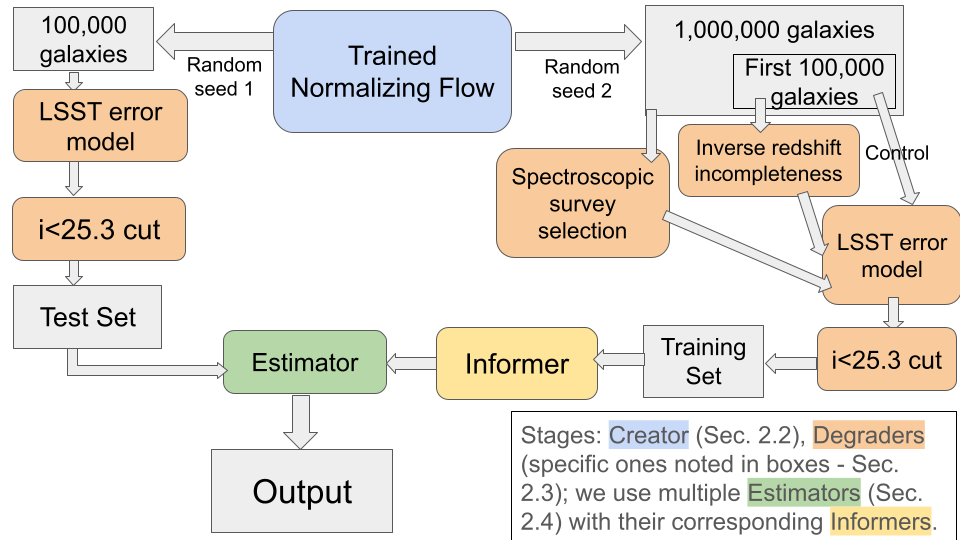}
  \caption{A flowchart depicting the pipeline structure used to perform creation, degradation, and estimation. \refresponse{Boxes with rounded corners are RAIL stages, while boxes with sharp corners are data products.}
  Creation \refresponse{stages} are shown in \cwr{blue}, degradation stages in peach, the informer stage \refresponse{that trains a given photo-$z$ estimation method} in yellow, and the estimator stage in green. \refresponse{The sections where more information may be found for the RAIL stages used are indicated in the legend, and Sec.~\ref{sec:overview} describes the overall workflow in detail.}}
  \label{fig:pipeline} 
\end{figure*}

\section{Methods \& Data} \label{sec:methodsdata}

Here we discuss the various features available in RAIL \refresponse{\citep{2025arXiv250502928T}} that have been employed in this work. 
In Section~\ref{sec:overview}, we summarize the pipeline structure used for our experiments.  We provide brief explanations of the simulated data in Section~\ref{sec:data} and our degradation methods in Section~\ref{sec:degraders}.
We provide visualizations of our degraded training data, and some relevant information on the surveys used to construct our spectroscopically degraded datasets. 
The photo-$z$ estimators we have examined are described in Section~\ref{sec:estimators}. 

\subsection{Overview of experimental design}
\label{sec:overview}






Our experiment employs RAIL-based  pipelines to efficiently examine a large number of test cases\refresponse{, with a workflow shown in Fig.~\ref{fig:pipeline}}. We employ RAIL's dataset creation and photo-$z$ estimation subpackages, and perform evaluation externally to RAIL. 

In our creation phase, first, training and test set samples were drawn from mock data generated with the PZFlow implementation of a normalizing flow \citep{Crenshaw24}. This normalizing flow was trained on the Rubin-Roman simulation data (see Sec.~\ref{sec:data}). The same test set was used in all cases across all estimators \refresponse{(left branch of Fig.~\ref{fig:pipeline})}, while a large number of different training sets were \refresponse{created as in the right-hand branch of that figure and examined to understand the impact of different forms of spectroscopic incompleteness}.

Our various training sets were obtained using different combinations of RAIL degradation stages. Our experiment \refresponse{was carried out using 3 types of training sets: training sets degraded with RAIL's inverse redshift incompleteness degrader, those degraded with spectroscopic selection from various surveys, and a control training set with neither form of degradation (designed to mimic the idealized case of a representative training set)}. \refresponse{For practical reasons, a larger initial pool of 1 million galaxies was needed when applying spectroscopic survey selections, as those severely downselect the galaxies, whereas the control and inverse redshift completeness training sets could be produced with a smaller initial population of $10^5$ galaxies.} 

\refresponse{As shown in Fig.~\ref{fig:pipeline}, after any other forms of degradation, all datasets  including the test set and control training set} were passed through RAIL's LSST error model degrader, \refresponse{applied with parameters corresponding to LSST 10 yr-depth \citep{ivezic2019}. The LSST error model stage interface through the Python package \texttt{PhotErr}, which is described in Appendix~B of \citep{Crenshaw24}.}
This introduces realistic LSST-like error bars and noise into our data. Finally, we perform an $i$-band magnitude cut on all data, to avoid training and testing on objects that would not be selected for cosmological analysis with the LSST. 

In our estimation stage, each of the five estimators listed in Section~\ref{sec:estimators} were trained \refresponse{(the ``Informer'' stage)} on each degraded training set and the control training set, and tested on our standardized test set.
After our introduction of noise and magnitude cuts, our resulting test set contained approximately 31,000 galaxies.
The training set size varied based on the degradation method, with our smallest catastrophically incomplete case containing only 195 objects, and our largest training set containing over 80,000. \refresponse{A typical photo-$z$ training set needs to have on the order of 10,000 training galaxies \citep[page 12 of ][]{Newman2015}, therefore, we warn that the results with a smaller training set will suffer from sampling noise.  In a real survey, the size of the training set is usually larger than this size, so the results of extremely small training might not show a realistic effect. }

In order to analyze estimation results, a variety of different performance metrics were employed, discussed in detail in Section~\ref{sec:metrics}. Performance metrics were implemented outside RAIL infrastructure, with \texttt{numpy} (\citealt{NumPy}), \texttt{scipy} (\citealt{SciPy}), or by hand. 


\subsection{Data and sample selection}
\label{sec:data}



\subsubsection{Truth catalog emulation}
\label{sec:truthcat}

Data from Tile 10307 from the Rubin-Roman simulation public data preview (\citealt{https://doi.org/10.26131/irsa569}; \citealt{2025MNRAS.544.3799O}) was used to train our normalizing flow. 
This data is reported in fluxes, as opposed to the magnitudes desired for this work, so all data was first converted to magnitudes using the standard SDSS definition of magnitude and LSST magnitude zero points. 
Tile 10307 contains approximately 3.5 million objects. 
Before training, all galaxies with \cwr{true} magnitudes fainter than 28 \cwr{in any band} were removed, leaving approximately 1.4 million objects. 
Of these remaining objects, 1 million were randomly selected and used to train the normalizing flow accessible through the RAIL creator stage employing PZFlow \citep{Crenshaw24}.

The normalizing flow is a sophisticated model for the joint probability density of redshift and magnitudes that we use as the basis for our mock data, which can be thought of as an interpolation of the seven-dimensional probability space of redshift and magnitude defined by the training set, which was the Roman-Rubin simulation in our case.
After generating a model with PZFlow's normalizing flow creator, initial training and test sets, comprised of true redshifts and magnitudes, were sampled. 
For our test set, 100,000 galaxies were sampled from the model, passed through the LSST error model with default parameters, and then subjected to an $i$-band magnitude cut of $i<25.3$. {We want to note that our flow models are trained on about 1.1 deg$^2$ of catalog, so we expect the truth catalog to have significant sample variance. } \refresponse{However, we do not expect the sample variance to significantly impact our conclusion. Future work is needed to propagate sample variance to the RAIL photo-$z$ results \citep{sanchez2020}.}


\subsubsection{True posterior PDFs}
\label{sec:data:true_pdf}

In this work, we compare the estimated Probability Density Function (PDF) of the photo-$z$ to the ``true PDF'' of a galaxy. 
Normally, the true galaxy redshift would be one number. 
However, the concept of ``true PDF'' is physically motivated since a collection of galaxies at different redshifts could yield the exact same magnitude due to galactic physics. 
Therefore, we define the true PDF of galaxy redshift as the conditional probability distribution of redshift given the magnitudes in each band. 
Statistically speaking, the true PDF of a galaxy with a set of magnitudes $m$ is
\begin{equation}
    P(z|m) = \frac{P(z, m)}{P(m)},
\end{equation}
where $P(z, m)$ is the joint probability density of the redshift-magnitude space, and $P(m)$ is the marginalized probability of the set of observed magnitudes $m$.  \cwr{In this work we are using a simulated dataset with some $P(z, m)$ that may differ in practice from realistic $P(z, m)$ due to uncertainties in galaxy populations and their evolution.  However, since we use the simulation for both training and testing set, any limitations in the simulated sample would apply to both and would not induce a training-testing set mismatch.}

We obtained the true posteriors for our test set using RAIL's \texttt{FlowPosterior} stage. 
With its creation capabilities, PZFlow yields true model posteriors by evaluating conditional probabilities for each galaxy from the same joint probability space of redshift-magnitude data from which the actual redshifts and magnitudes of the test set were drawn. 
We provided PZFlow with our test set magnitudes, and used this functionality to generate our true redshift posteriors.;
as \texttt{FlowPosterior} generates unique conditional probabilities for each galaxy (as opposed to one single distribution applied to all galaxies), it obtains the true posterior for each galaxy from the cross-section of the conditional probability distribution $P(z|u_i,g_i,..., y_i)$ corresponding to its specific photometry. 

\subsection{Training set degradation}
\label{sec:degraders}



RAIL's degradation algorithms are used to introduce systematic error into training sets of photometric data. 
Differences between training data and test data, such as training set incompleteness and non-representativity, affect photo-$z$ algorithm performance, but the specific training set conditions to which various estimation algorithms are sensitive or performant is not yet adequately understood. 
In this work, we \cwr{begin with a simple case of training set incompleteness at high redshifts in Section~\ref{sec:invz}.  We have next chosen} to examine the effects of training set degradation based on data collected by previous surveys, detailed in Section~\ref{sec:survs}, as existing survey data will likely be used in future science involving photo-$z$ estimation. 
 In all of these cases, we assume LSST-like photometric noise as described in Section~\ref{sec:lssterrmod}. \refresponsetwo{Note that in a real spectroscopic survey, the selections are done on photometry with noise from the corresponding imaging survey for the target selection. In this work, we ignored this photometric noise and assumed that the spectroscopic selection is done on perfect photometry, then applied the photometric noise degrader.} 

\begin{figure*}
  \centering
  \includegraphics[width=1.0\textwidth]{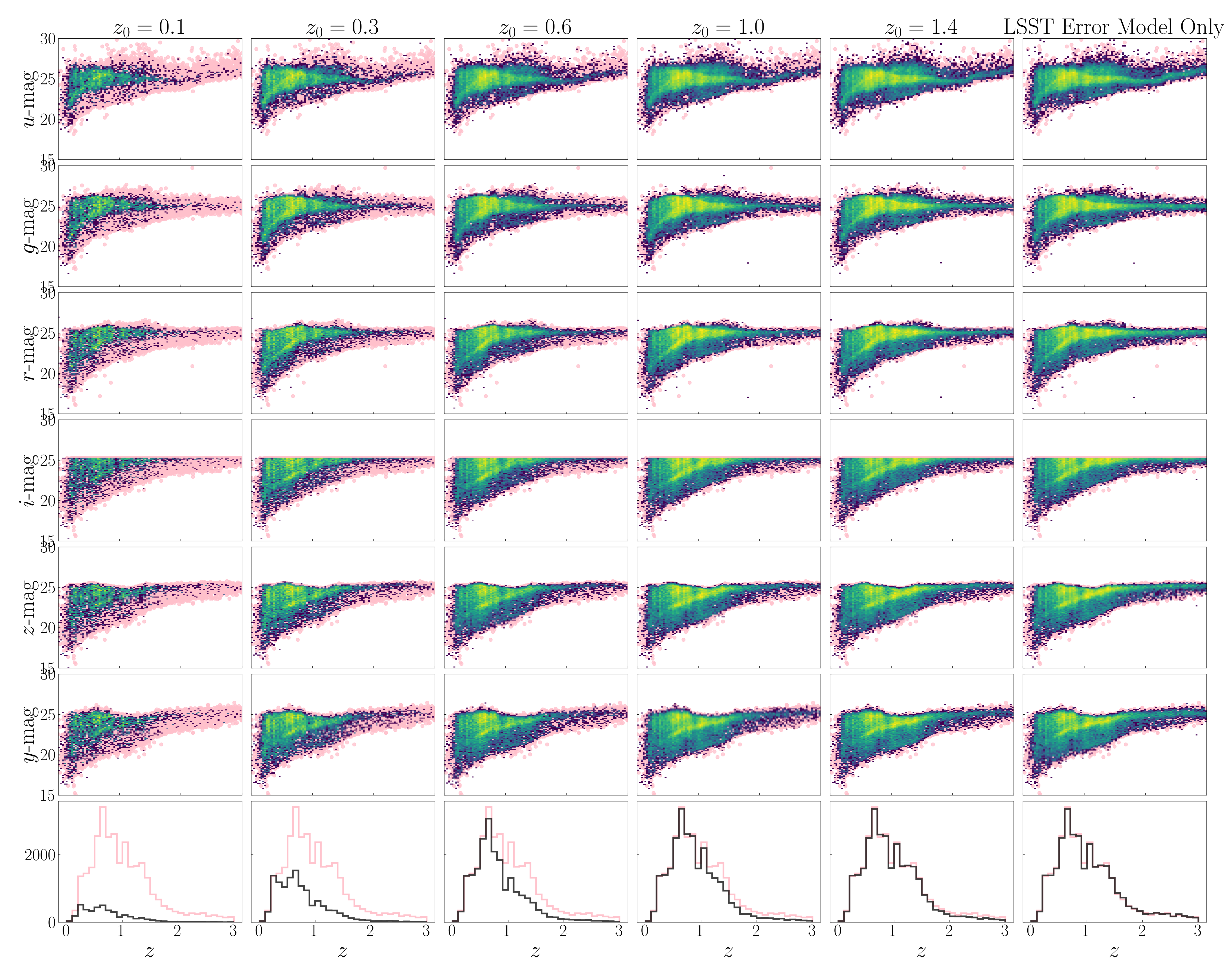}
  \caption{Regions in redshift-magnitude space  remaining populated after inverse redshift incompleteness degradation, performed with several different values of $z_0$ (as labeled at the top of each column). 
  The colormap indicates the density of surviving data, while the pink points have no inverse redshift incompleteness degradation and are our test set. \refresponse{The top 6 rows show the 6 LSST passbands, while the bottom row shows the 1D histograms of redshift before and after the degradation.} 
  }
  \label{fig:invz trainsets} 
\end{figure*}

\subsubsection{Inverse Redshift Incompleteness}
\label{sec:invz}

First introduced in \citet{Stylianou22},  \refresponse{the inverse redshift incompleteness degrader can be used to generate semi-realistic spectroscopic survey success rate which has lower completeness after a pivot redshift, either from fainter magnitude of the high-redshift galaxies, or that the spectroscopic features fall outside of the spectrograph’s wavelength coverage. } Inverse redshift incompleteness degradation is implemented as a function of some pivot redshift value $z_0$. 
Each galaxy in a dataset is assigned a survival probability as a function of this value, given by $1 - \text{min}(1, \frac{z_0}{z})$, where $z$ is the galaxy's redshift. 
Galaxies are then removed from a dataset according to this probability distribution, with the intent of replicating the incompleteness at higher redshifts seen in training data from real sky surveys. 
As \citet{Stylianou22} was completed before RAIL was released, we test the RAIL implementation with the same pivot values to confirm that we can recover results consistent with the original implementation, demonstrating the validity of our pipeline in the one case previously explored in the literature. We also include some higher values of pivot redshift in our work, in order to recover trends in performance as $z_0$ is changed. 
We test pivot values of $z_0 = \{0.1, 0.3, 0.6, 1.0, 1.4$\}. The impact of this degrader on the magnitude versus redshift relation is shown in Fig.~\ref{fig:invz trainsets}. \refresponse{The inverse redshift incompleteness skew the overall redshift distribution with different $z_0$ values. The $z_0 = 0.1$ model selects few objects overall due to an extremely low pivot redshift, although this is an extreme scenario. }

\begin{figure*}
  \centering
  \includegraphics[width=1.0\textwidth]{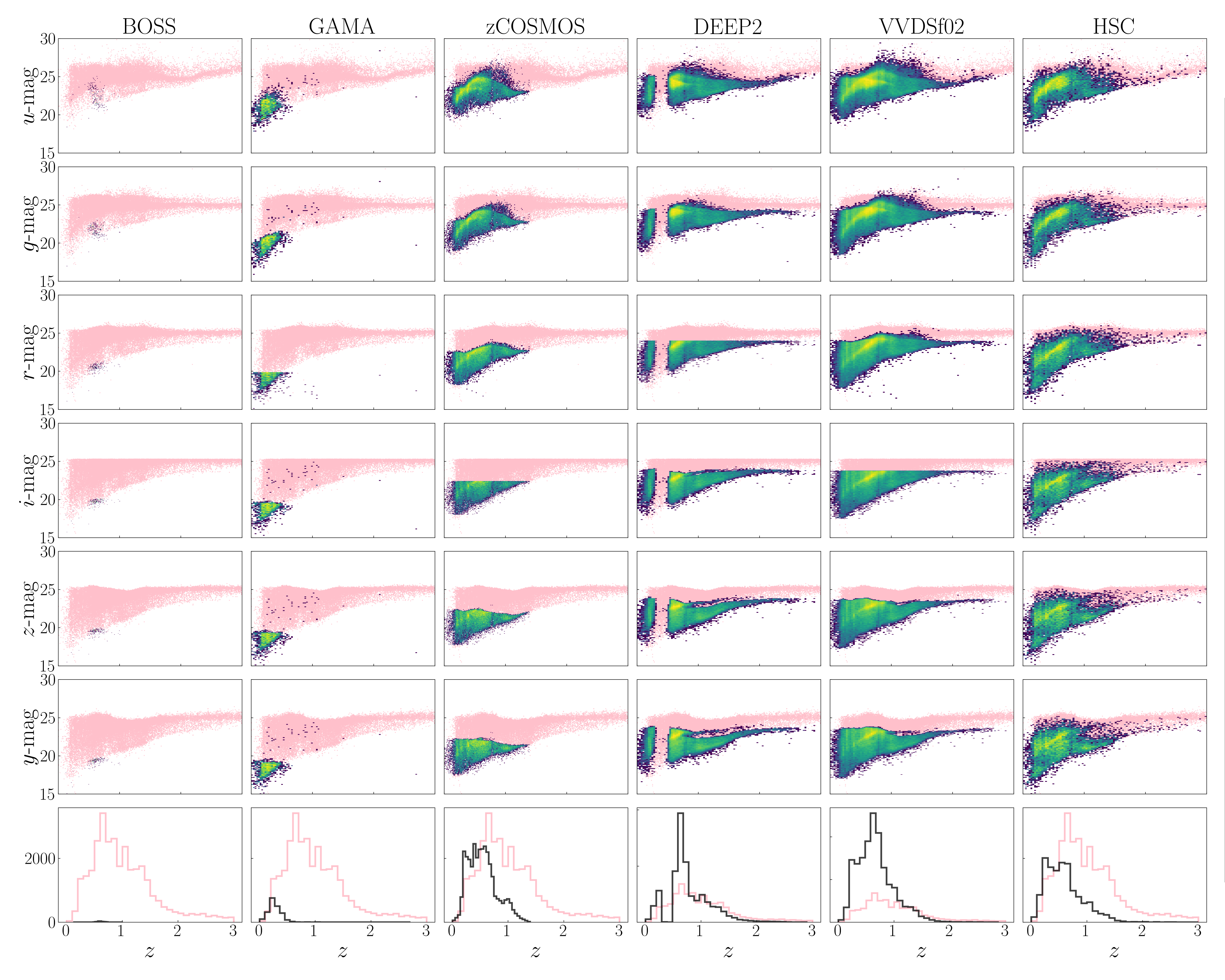}
  \caption{Regions in redshift-magnitude space  remaining populated after spectroscopic survey degradation. 
  The colormap indicates the density of surviving data, while the pink data have received no spectroscopic survey degradation and are our test set (see Fig.~\ref{fig:pipeline}). Note the variation in degrees of test set coverage and localization of data in redshift-magnitude space across different surveys. \refresponse{The top 6 rows show the 6 LSST passbands, while the bottom row shows the 1D histograms of redshift before and after the degradation.}}
  \label{fig:spec trainsets} 
\end{figure*}

\subsubsection{Survey-Based Degradation}
\label{sec:survs}

RAIL contains degradation stages designed to mimic the incompleteness and non-representativity patterns in spectroscopic data collected or used by past surveys. 
We employed six such degraders: 
\begin{enumerate}
\item The Sloan Digital Sky Survey (SDSS) Baryon Oscillation Spectroscopic Survey (BOSS, \citealt{2013AJ....145...10D}): 
We use the target selection described in \cite{2016MNRAS.455.1553R}.  
BOSS is a wide-area survey targeting relatively bright, typically red, galaxies for the purpose of measuring large-scale structure rather than targeting a flux-limited sample for training photometric redshift methods, so it is particularly non-representative of anticipated LSST samples. 
Its spectroscopic success rate is very high, so the main impact of this degrader is to apply the targeting criteria to the photometry.\footnote{Caveat: Very few galaxies in our training data survive BOSS degradation, due to these heavy color and magnitude cuts. Due to its tiny size, one would expect estimates made using only BOSS training data to perform even more poorly by our metrics than those made with GAMA training data, so it is included in our analysis largely to contrast with GAMA. We recognize that both surveys result in catastrophically and unrealistically incomplete training data, however the ways in which some estimators respond to the two cases were surprising.} 
%
\refresponse{Since our degrader implements an approximate version of the BOSS target selection, we refer to this case hereafter as `pseudo-BOSS'. }
\item The Galaxy And Mass Assembly (GAMA, \citealt{2022MNRAS.513..439D}) survey: 
This survey observed a relatively bright sample of galaxies (compared to LSST samples) over a few hundred square degrees.  
For this survey, we use the approximate 90\% completeness limit of $r<19.87$ and do not apply any spectroscopic incompleteness for the selected galaxies. 
\item The Hyper Suprime-Cam (HSC) survey \citep{2018PASJ...70S...4A}:
    Unlike the other degraders, this one is associated with a photometric survey (HSC), and corresponds to the merged spectroscopic sample used by the HSC team to train photometric redshifts.  It was implemented in \citet{2023ApJ...950...49M} as the union of multiple different spectroscopic surveys\footnote{For details of quality cuts imposed on the samples, please see \url{https://hsc-release.mtk.nao.ac.jp/doc/index.php/catalog-of-spectroscopic-redshifts__pdr3/}}, as described in \citet{2020arXiv200301511N}.  
    These include all of the above-mentioned samples for which there are individual degraders, though the HSC GAMA dataset is from an earlier release, data release 2 \citep{2015MNRAS.452.2087L}, and the HSC SDSS dataset is from a later release that includes the extended Baryon Oscillation Spectroscopic Survey \citep{2020ApJS..249....3A}.  
    It also includes the following additional samples: UDSz\footnote{\url{https://www.nottingham.ac.uk/astronomy/UDS/UDSz/}} \citep{2013MNRAS.433..194B,2013MNRAS.428.1088M}; 3D-HST\footnote{\url{https://archive.stsci.edu/prepds/3d-hst/}} \citep{2014ApJS..214...24S,2016ApJS..225...27M}; FMOS-COSMOS\footnote{\url{https://member.ipmu.jp/fmos-cosmos/FC_catalogs.html}} \citep{2015ApJS..220...12S}; VIPERS\footnote{\url{http://vipers.inaf.it/}} public data release 1 (PDR1; \citealt{2014A&A...562A..23G}); SDSS quasar sample from data release 14 \citep{2018A&A...613A..51P}; WiggleZ\footnote{\url{https://wigglez.swin.edu.au/site/forward.html}} data release 1 \citep{2010MNRAS.401.1429D}; DEEP3\footnote{\url{https://sites.uci.edu/deep3/}} \citep{2011ApJS..193...14C,2012MNRAS.419.3018C}; PRIMUS\footnote{\url{https://primus.ucsd.edu/}} data release 1 \citep{2011ApJ...741....8C,2013ApJ...767..118C}; the 2dF Galaxy Redshift Survey (2dFGRS\footnote{\url{http://www.2dfgrs.net/}}; \citealt{2003astro.ph..6581C}); the 6dF Galaxy Redshift Survey (6dFGRS\footnote{\url{http://www.6dfgs.net/}}; \citealt{2004MNRAS.355..747J,2009MNRAS.399..683J}); the Complete Calibration of the Color-Redshift Relation (C3R2\footnote{\url{https://sites.google.com/view/c3r2-survey/home}}) survey \citep{2017ApJ...841..111M,2019ApJ...877...81M}; the DEIMOS 10k sample \citep{2018ApJ...858...77H}; the Large Early Galaxy Census (LEGA-C\footnote{\url{https://www.eso.org/sci/publications/announcements/sciann17120.html}}) survey data release 2 \citep{2018ApJS..239...27S}; and VANDELS\footnote{\url{http://vandels.inaf.it/}} data release 1 \citep{2018A&A...616A.174P}. 
    In practice, this heterogeneous spectroscopic degrader applies both a cut on the photometry to reflect the targeting criteria of the spectroscopic datasets, and downsampling to reflect the spectroscopic success rate.  
    Both cuts are done in bins in the 2D space of $g-z$ color and $i$-band magnitude. 
 \item The zCOSMOS survey: 
    We use the spectroscopic redshift success rate as a function of redshift and $i_\text{AB}$-band magnitude for the zCOSMOS-bright survey ($i_\text{AB}<$22.5) from the first two years of observations \citep[figure 3 of][]{2009ApJS..184..218L}.
\item The DEEP2 galaxy redshift survey \citep{2013ApJS..208....5N}: 
This survey covers 4 fields totaling 2.8 deg$^2$ down to $R_\text{AB}=24.1$.  
There are additional color cuts in three fields to target galaxy samples with $z \gtrsim 0.7$.  
While not precisely correct, the cuts on $B-R$ and $R-I$ in the DEEP2 targeting photometry are applied instead to $g-r$ and $r-i$ in our mock LSST dataset.  
The spectroscopic success rate is applied as a function of magnitude based on \citet{2013ApJS..208....5N}\footnote{Caveat: According to private correspondence with members of the DEEP2 team, the publicly available cut values may not be those that actually correspond to the survey data; as an official correction has not yet been issued, we nonetheless proceed with the posted values. \refresponse{We refer to our dataset implementing these cuts as 'pseudo-DEEP2' in order to avoid potential confusion.} 
}.
   %
    \item The VIMOS VLT Deep Survey (VVDS; \citealt{2013A&A...559A..14L} for the final release of the full survey): 
    We use the spectroscopic redshift success rate as a function of $i_\text{AB}$ in the range $[17, 24]$ from Figure~16 in \cite{2005A&A...439..845L} for the VVDS 2h field.  
    The redshift success rate is applied independently rather than jointly with magnitude, which results in a pessimistic view. This degrader results in a sample that is like some combination of the cuts for VVDS bright and VVDS deep rather than either alone due to $z$-dependent success rate.
\end{enumerate}


{We emphasize that our spectroscopic degraders are designed to approximate the photometric distribution of the target selection of spectroscopic surveys. The real spectroscopic success rate also depends on other galaxy properties such as star formation rate and stellar mass, which are not considered in this work. }

Referring to Figure \ref{fig:spec trainsets}, we see an illustration of the distributions of the corresponding surveys. Real spectroscopic success also depends on the physical properties of the galaxy, such as data from each of these surveys in redshift-magnitude space. The pseudo-BOSS survey represents a catastrophically poor training set case; it is both severely incomplete and not representative of the true distribution of galaxies in redshift-magnitude space. Though GAMA data is denser in the regions it covers, it also represents a very incomplete portion of our test data, with the vast majority of data concentrated below $z\approx0.7$ and a magnitude of 20. zCOSMOS is very complete for redshifts less than $z\approx1.5$, but contains no data for higher redshifts, and also lacks magnitude depth. pseudo-DEEP2 has good coverage of much of the test dataset, but has a large gap between $z\sim0.3$ and $z\sim0.6$, and less depth in magnitude. VVDSf02 and HSC are the most complete of the datasets examined here. The VVDSf02 training set contains the most data points, and they populate much of the redshift-magnitude space in our test set, though depth is slightly lacking in some bands. Though less dense than VVDSf02, HSC is the most representative of the surveys we examined, with data populating nearly all of the redshift-magnitude space covered by our test set.

The variety of different cuts described above produce significantly different post-degradation training set sizes (listed in leftmost column of Table \ref{tab:spec rej}), as stricter cuts cause more galaxies to be removed. The several surveys examined here provide us not only with variation in color and magnitude cuts, but also with significant variation in training set size, which is also a relevant form of imperfection. Our training set sizes vary from $\sim$200 objects (pseudo-BOSS) at the smallest to $\sim$80,000 objects (VVDSf02) at the largest. Though we recognize that training sets smaller than $\sim$10,000 objects (pseudo-BOSS and GAMA) are likely catastrophically and unrealistically incomplete, we examine both cases here, as some algorithms do not treat these two cases as we would expect. 

\begin{figure}
\begin{center}
  \centering
  \includegraphics[width=0.5\textwidth]{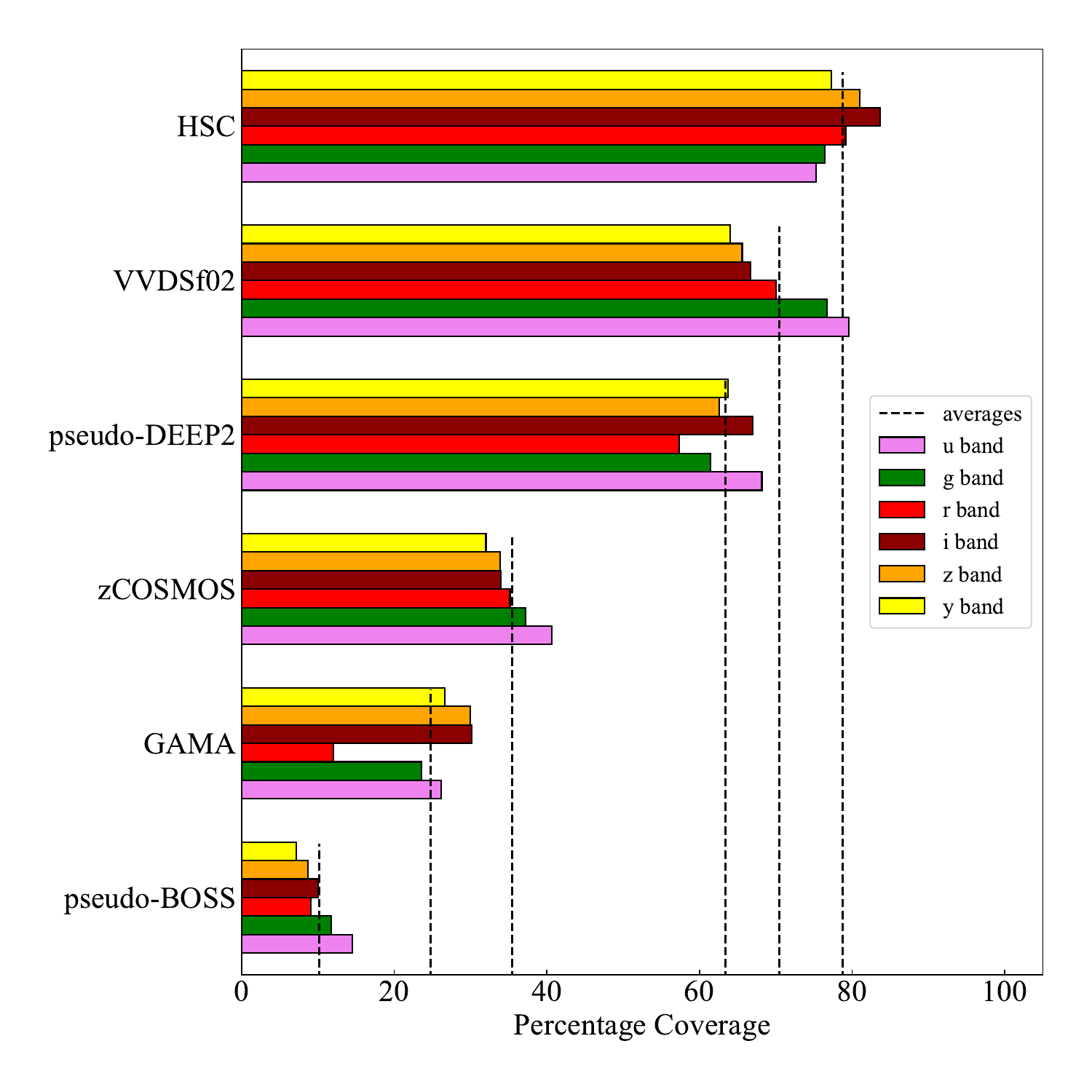}
\caption{Bars correspond to approximate percentages of our test set covered by each spectroscopic training set, shown for each LSST band. The range of redshift-magnitude space captured in each band was covered with a discretized grid, with bins of width 0.1 in the redshift direction and 1 in the magnitude direction. The reported percentage is the percentage of the squares containing at least one test set galaxy that also contain at least one training set galaxy. Dashed lines correspond to the average coverage across all six bands for each spectroscopic degrader.}

\label{fig:spec coverage} 
\end{center}
\end{figure}

Though training set size is certainly a helpful form of imperfection to consider, it is also important to consider the distribution of training data in redshift-magnitude space relative to test data. For example, though our zCOSMOS case contains significantly more data than our HSC case, zCOSMOS data is densely isolated in a relatively small region of the redshift-magnitude space, whereas HSC data is spread over nearly all of the test set. 
Figure~\ref{fig:spec coverage} shows our attempt to quantify test set coverage under each type of survey degradation. Colored bars show the percentage of each band in our test set covered by each training set, while dashed lines show the average across all six bands for each survey training set. 

\refresponse{We also note that a realistic training set for LSST will likely have spectroscopic and other reference samples from various sources matched to the photometric dataset, more like the HSC selection here and with of order $10^5$ galaxies.  Therefore, we do not expect the impact of the spectroscopic selection in a real survey to follow one of the single-survey selections, but rather the aggregate effect of all selections. }

\subsubsection{LSST Error Model}
\label{sec:lssterrmod}


Every training set used in our analysis was passed through the LSST error model degradation stage (section 3.2 of \citealt{ivezic2019}) after other degradation was applied, in order to approximate LSST-like photometric noise in training sets. Default parameters were used, corresponding to 10-year depth. 

We also constructed a control training set using the LSST-like error as our only form of degradation. 
Every algorithm was trained once on a dataset designed to be representative of its test set, in order to provide a ``baseline'' for ideal training set conditions. 
Before making magnitude cuts, this data was degraded with the LSST error model only, and sampled from the same larger set of galaxies as the test set, using a different random seed. 


\subsection{Estimators}
\label{sec:estimators}


This study is not intended to be a comprehensive data challenge between all estimators under consideration for LSST;
rather, it is intended to examine how we can best assess the relative sensitivity of estimators as training set imperfection worsens, so we understand what test cases are meaningful and how to interpret assessment metrics in a full-scale, more realistic experiment with all estimators and their tuning parameters.
As a result, we have chosen a subset of photo-$z$ estimators representing categories of relevance\refresponse{:}
%
%
\begin{enumerate}
    \item 
Our experiment uses \texttt{trainZ}, first introduced in \citeDCt, as an algorithmic experimental control. 
It assigns each test set galaxy a photo-$z$ PDF equal to the redshift distribution of the training set, without taking into account the photometry of the galaxies in the test set. 
{\texttt{trainZ} performs poorly in point estimate metrics and CDE loss}, yet it outperformed the real estimators by nearly all metrics in \citeDCt due to {the fact that the training set was representative of the test set. In this work, we expect \texttt{trainZ} to perform poorly due to the non-representative training set we employ. }
%
%
\item \texttt{FlexZBoost} is a conditional density estimator introduced in \citet{Izbicki17} and demonstrated in the context of astronomy in \citet{Dalmasso20}.
It uses the distribution of training set galaxies close to each test set galaxy in color-space to approximate the distribution of redshifts of galaxies using a Fourier basis.
It is included in this study because it performed best by the only metric in \citeDCt that appropriately penalized \texttt{trainZ} and is thus likely to be used for LSST.
%
%
\item The Color-Matched Nearest Neighbor algorithm, implemented as \texttt{CMNN}, was first introduced in \citet{Graham18}.
It is included in this study because it is used throughout the \textsc{LSST} ecosystem for testing and optimization purposes, for example, for assessing the impact of observing strategy on photo-$z$ quality.
For each test set galaxy, \texttt{CMNN} identifies a subset of training set galaxies with similar colors and assigns the test set galaxy a photo-$z$ PDF equal to a Gaussian approximation to the redshift distribution of that set of galaxies;
for this reason, we will refer to it as a ``Gaussian estimator.''
\texttt{CMNN} differs from the standard k-nearest neighbor algorithm in that the subset is not defined as the fixed number k of nearest neighbors in the space of photometry.
Rather, the subset is defined by a distance that accounts for the observational errors of the test set galaxy, so the number of neighbors depends on the density of points in the training set as well as the precision of the test set galaxy's photometry. 
%
%
\item Gaussian Process photo-$z$, implemented as \texttt{GPz}, models the space of photometry and redshift with a Gaussian process, as described in \citet{Almosallam15, Almosallam16}.
While the most up-to-date \texttt{GPz} applys a Gaussian process to the training set and evaluate a posterior PDF of a test set galaxy on a grid based on the same Gaussian process, \texttt{GPz} implemented in RAIL currently assumes the individual test set photo-$z$ PDFs are Gaussian distributions whose variances contain contributions from both aleatoric uncertainty due to the training set's sparsity and epistemic uncertainty due to the test set photometric errors;
we will also refer to \texttt{GPz} as a ``Gaussian estimator.''
\texttt{GPz} is included here to enable consistency checks with \citet{Stylianou22}, the only previous study to consider any of our test cases.
%
%
\item \texttt{pzFlow} \citep{Crenshaw24} models the joint probability distribution of photometry and redshift with a normalizing flow, an effective interpolation scheme of the seven-dimensional space.
Photo-$z$ PDFs are then evaluated as posteriors, at fixed photometry and evaluated on a grid of redshifts.
\texttt{pzFlow} is included in this study for self-consistency checks against the forward modeling approach, as the same underlying algorithm serves as the back-end for creating the mock data used for all test cases.
\end{enumerate}

\section{Performance Metrics}
\label{sec:metrics}




Various performance metrics 
were used to examine the accuracy of the outputs of each algorithm under the chosen training set conditions. 
The use of several different performance metrics is important in our analysis in order to obtain a holistic understanding of algorithm sensitivity, as some metrics may not be able to detect the impact of imperfect training data 
while others could potentially flag opportunities for algorithmic improvement in this area.
All performance metrics were implemented with \texttt{scipy} \citep{SciPy} or \texttt{sklearn} \citep{scikit-learn}, or constructed by hand outside of RAIL infrastructure\footnote{Some of these are available in RAIL v1, a later version of the software than was used in this work.}. 

 Distribution-to-distribution metrics, described in Section~\ref{sec:distdist}, were used to compare estimator outputs \revone{}{$\hat{p}(z | x_{i}, \pi_{j})$ of redshift $z$ given the $i$th galaxy's photometry $x_{i}$ for each estimator $j$'s implicit prior $\pi_{j}$} to the test set true posteriors\revone{, and point-to-distribution}{ $p(z | x_{i})$.
\refresponsetwo{CDF-based metrics, described in Section~\ref{subsec:cdf}, were used similarly, but for cumulative distribution functions.}  Point-to-distribution
}
metrics, described in Section~\ref{sec:pointdist}, were used to compare to true (spectroscopic) redshifts also obtained from the test set. 
Mode point estimates \revone{}{$\hat{z}_{i, j}$} obtained from estimator outputs \revone{}{$\hat{p}(z | x_{i}, \pi_{j})$} were also compared to the\revone{se}{ true (}spectroscopic\revone{}{)} redshifts \revone{}{$z_{i}$} using the metrics of Section~\ref{sec:outlier rejection}. 


\refresponse{In most cases, the detailed definition and implementation of the metric is described in Appendix~\ref{app:metrics}, and we summarize here the key aspects of the metric needed to understand the results.}

\subsection{Distribution-to-Distribution Metrics}
\label{sec:distdist}



Since we are able to generate true posterior distributions for our test data (see Section \ref{sec:truthcat}), it is valuable to compare \revone{}{each estimator's} output distributions to these true posteriors. 
Comparing \revone{distributions}{probability densities} to \revone{distributions}{probability densities} gives us the most complete picture of true algorithm performance, as whole distributions encode significantly more information than point estimates do. 
Though in some instances we may use point redshifts as our ``truth'' (see Section \ref{sec:pointdist}\revone{}{)}, under cases of realistic complexity it is rare that we will have access to spectroscopic point redshifts for all of our test data. 
\revone{It is also valuable to employ \texttt{RAIL}'s \texttt{FlowPosterior} stage in our work, as this new capability has not been explored much in previous work.}{}

Both PDFs and \revone{}{their cumulative distribution functions (}CDFs\revone{}{)} may be obtained easily and efficiently from array-like statistical ensembles with the Python library \texttt{qp} \citep{malz_approximating_2018}.
\revone{}{It was also valuable to put into use \texttt{RAIL}'s \texttt{FlowPosterior} stage in our work, as this new capability has not been explored much in previous work.} \refresponsetwo{The distribution-to-distribution metrics we use are as follows:}

\refresponse{
\begin{itemize}
\item Root Mean Square Error (RMSE): The Root Mean Square Error, or Root Mean Square Deviation, is the quadratic mean of a set of differences between true values and predicted values. \refresponsetwo{As such, it is particularly sensitive to outliers.}
\item Kullback-Leibler Divergence (KLD): The KLD is another statistical \revone{distance}{measure of discrepancy}, used to measure the \revone{difference}{divergence} between a reference (true) probability distribution and \revone{a second}{an approximating} (estimated) distribution\revone{}{, in the form of the information lost by using the approximation instead of the truth}. 
\end{itemize}
}

\subsection{CDF-based metrics}\label{subsec:cdf}
     

\revone{}{In our descriptions of metrics based on the CDF, the integral of a PDF, we denote the CDF of each galaxy $i$'s true and estimated posterior as $F_i(z)$ and $f_i(z)$ respectively.
When the CDF is estimated from samples, it is technically an empirical CDF (ECDF), which we denote as $\hat{F}(z)$ if it is derived from the true redshift posteriors and $\hat{f}_{j}(z)$ if derived from the $j$th estimator's redshift posteriors.
Both CDFs and ECDFs are obtained from the statistical ensembles output by RAIL stages using \texttt{qp} \citep{malz_approximating_2018}.} \refresponsetwo{The CDF-based metrics we use are as follows:}

\refresponse{
\begin{itemize}
\item Kolmogorov-Smirnov \revone{}{(KS)} Test Statistic: The KS defines a \refresponsetwo{maximum} distance between a \revone{cumulative distribution function}{CDF}, in our case \revone{the CDF}{that} of the true posterior generated for the test data, and an \revone{empirical distribution function}{ECDF}, in our case an estimator's output. \refresponsetwo{It is particularly sensitive to the central part of the distribution.}
\item Cramér-von Mises \revone{Criterion}{} (CvM): The Cramér-von Mises criterion is an alternative to the KS Test\revone{,}{ Statistic} in which the mean square distance between distribution functions is used\refresponsetwo{, potentially introducing sensitivity to other parts of the distribution beyond the center compared to the KS test}.
\item Earth-Mover's Distance (First Order Wasserstein Distance): The Wasserstein Distance is a measure of the distance-weighted integrated probability density of one distribution with respect to another distribution.
\end{itemize}
}

\subsection{Distribution-to-Point Metrics}
\label{sec:pointdist}

Distribution-to-point metrics may be employed when true galactic redshifts are known to high enough precision that point estimates are appropriate, or in the absence of true posteriors with which to calculate distribution-to-distribution statistics. 
We examine two distribution-to-point metrics commonly referenced in previous literature. 

\refresponse{
\begin{itemize}
\item Probability Integral Transform (PIT): A statistic that quantifies whether the per-galaxy Bayesian posteriors are consistent with the corresponding population of true redshift values.
%
    We expect the PIT distribution for more representative and complete training sets to more closely resemble a uniform distribution. 
    Common features of non-representative PIT distributions are ``side spikes'' and ``middle bumps''\revone{: ``s}{.
    ``S}ide spikes'' indicate an over-abundance of instances where the truth is entirely on one side or the other of an output distribution, indicating that either \revone{estimates are wildly incorrect}{the bulk of the probability density is far from the true value} or that output distributions are too narrow. 
    ``Middle bumps'' indicate that the truth divides the output probability density roughly in half too often, indicating that an estimator may be over-fitting to training data or that output distributions are too \revone{narrow}{broad}. 
    PIT distributions are compared to the uniform distribution using all the metrics of Section~\ref{sec:distdist}.
\item Conditional Density Estimate (CDE) Loss: The CDE Loss is an extension of the RMSE for PDFs that evaluates expectation values of the posterior PDFs of redshift conditioned on photometric data\refresponsetwo{, and is sensitive to outliers, scatter, and to a lesser extent biases in photo-$z$}. 
\end{itemize}
}

\begin{figure*}
  \centering
  \includegraphics[width=0.8\textwidth]{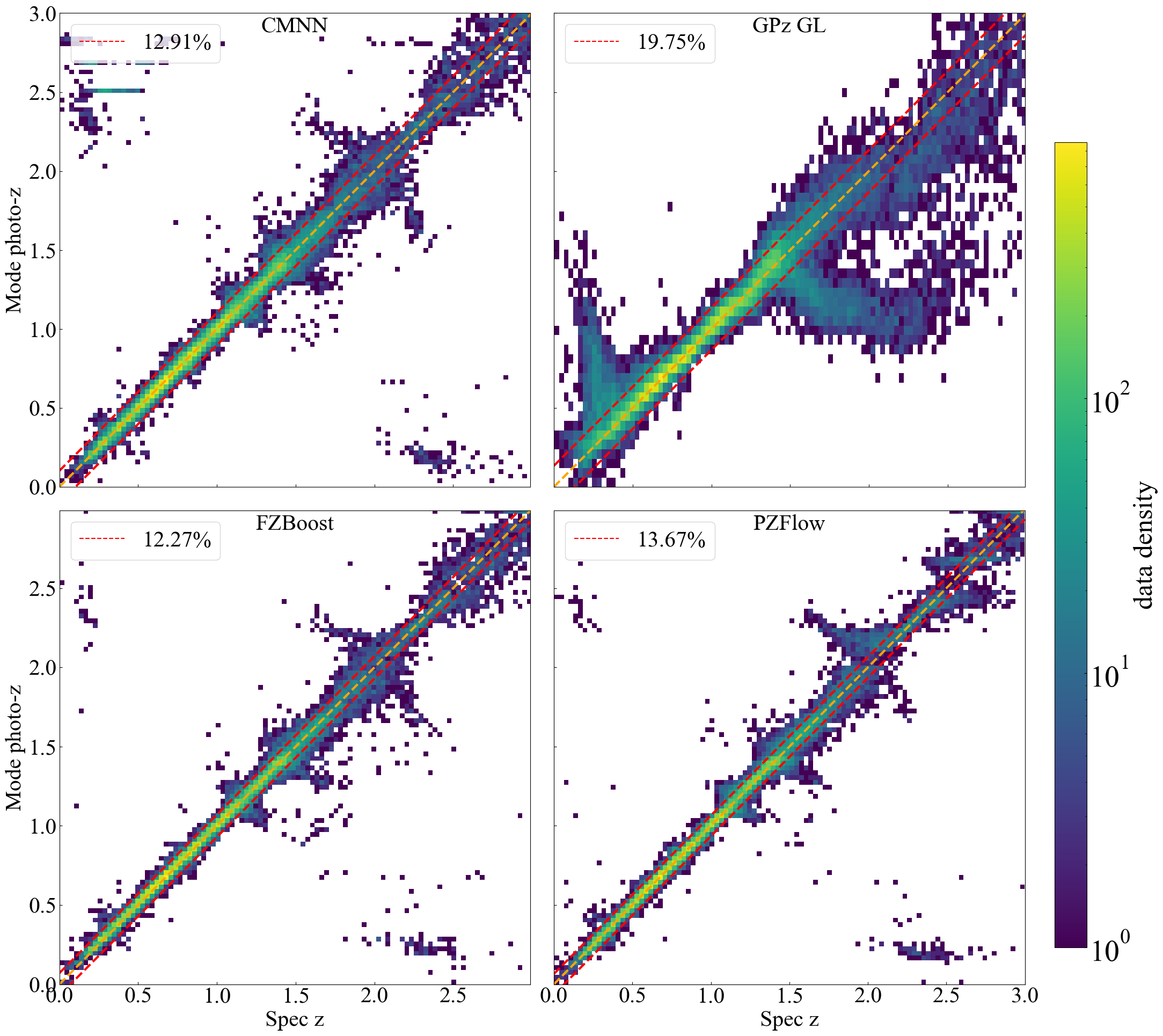}
  \caption{Mode point estimates obtained from estimator outputs under representative training conditions, versus true spectroscopic galaxy redshifts. Clustering along the diagonal is ideal case, which is shown by the dashed orange line. Red dashed lines show the 3$\sigma$ distance, calculated via outlier rejection. \texttt{TrainZ} is not shown, as the plot is not illustrative: it assigns every galaxy the same distribution, thus all modes are the same. $>3\sigma$ outlier rates are shown in the upper left corner of each panel. 
  }

  \label{fig:control modes} 
\end{figure*}

\subsection{\revone{Ensemble}{Point-to-point} Metrics}
\label{sec:pointestmets}

    \refresponsetwo{One approach to quantitatively characterizing catastrophic outliers in the photo-$z$ point estimates is to statistically identify outliers and exclude the flagged outliers from further analysis.  Here we use the following algorithm:} 
    For a set of numbers, its mean and standard deviation are calculated. 
    Any elements greater than 3 standard deviations from the mean of the dataset are then \revone{labelled}{labeled} as catastrophic outliers and `rejected'. 
    The mean and standard deviation of the remaining dataset is then recalculated, and the process repeated recursively until convergence is reached and no more points are being rejected. 
    The rejected points are the catastrophic outliers of the dataset.
    We have implemented this metric by hand. 
    
    We use outlier \revone{rejection}{identification} to assess the mode point estimates obtained from estimator output PDFs. 
    For each data point in the plots of photo-$z$ mode versus spec-$z$ (for example, see Figure~\ref{fig:control modes}), its distance from the diagonal was calculated, and outlier \revone{rejection}{identification} was then performed on this set of distances in order to obtain a catastrophic outlier rate. 
    Though this metric is helpful in quantifying cases of extreme error, it is important to note that it can be easily biased by datasets with very large standard deviations, and the result will be artificially low in such cases. 
    \texttt{trainZ} is a prime example of this; as every point estimate for \texttt{trainZ} is the same, the standard deviation of the set of their distances from the diagonal is very large, thus nearly all points fall within $3\sigma$ of the diagonal. 
    We see this also in some cases of catastrophically poor estimation. 
    \label{sec:outlier rejection}




\begin{figure*}
  \centering
  \includegraphics[width=0.65\textwidth]{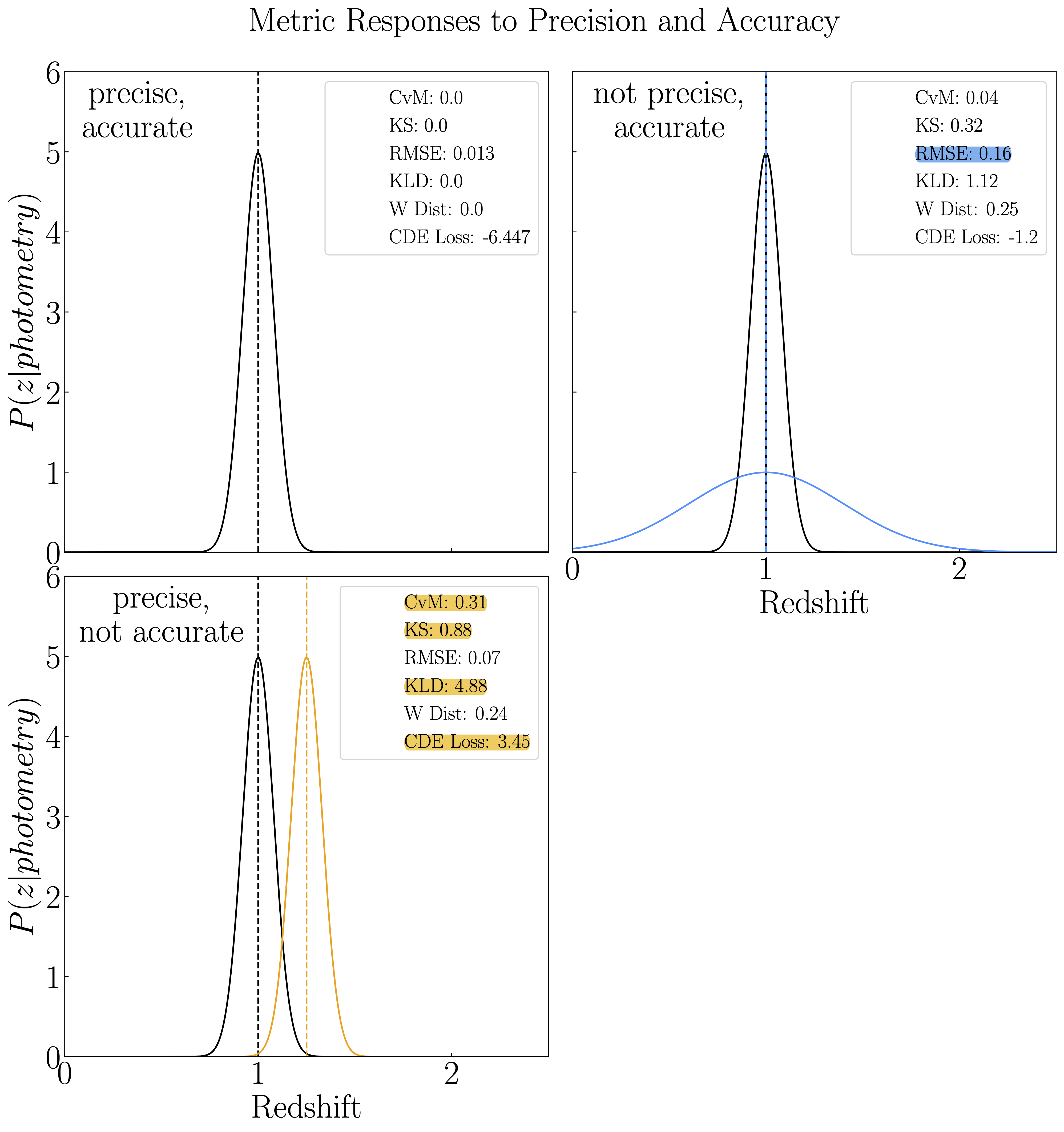}
\caption{Illustration of distribution-to-distribution metric responses to bias on the centroid and width of the distribution. 
Each panel shows the reference redshift distribution (black) alongside an altered version, representing scenarios with increasing scatter (top right), and biased centroid (bottom left). 
The numerical values of these statistics are shown in each panel to illustrate how they respond to different types of distributional changes. \cwr{The Wasserstein distance responds similarly to imprecision as to inaccuracy. The highlighted metrics in the legend indicate the cases where a metric is more sensitive to inaccuracy or to imprecision.} }

  \label{fig:stdev fx} 
\end{figure*}

\subsection{Understanding the Metrics}
\label{sec:distshape}


\revone{}{In Figure~\ref{fig:stdev fx}, we illustrate how the distribution-based metrics react to two types of mischaracterization of the true PDF -- (a) overall bias in the distribution, (b) difference in the width of the distribution. }

\revone{}{The top left panel of Figure~\ref{fig:stdev fx} shows the distribution-based metrics comparing the PDF to itself. In this case, the CvM, KS, KLD, and Wasserstein distance are exactly zero. The RMSE is approximately $2 {\rm Var}[X]$, as predicted by its definition, which in this case is $2\times (0.08^2) = 0.0128$. The CDELoss is a negative value. }

\refresponse{Moving to the bottom left panel, w}e see that \revone{}{bias} between distributions is penalized heavily in the CvM, KS, KLD and CDELoss. 
It is penalized less in the RMSE, which is expected given the findings in \citet{malz_approximating_2018} \revone{which}{that} indicate relative insensitivity of the RMSE to distribution tails when compared with the KLD. 
Conversely in the KS test, similarity in shape rather than overlap results in a lower value\revone{}{, which makes sense since it is sensitive to the largest discrepancy in ECDF.  In the inaccurate case, that will just be a constant shift related to the area under one distribution where the other has zero probability, whereas in the imprecise case, it will depend on the slope of the CDF that comes from how much broader the approximating distribution is}. 

\revone{}{Overall, \cwr{the illustrated cases of inaccuracy and imprecision result in roughly the same Wasserstein distance, while the RMSE is more sensitive to an incorrect width of the distribution (top right panel), and the CvM, KS, KLD and CDELoss are more sensitive to biases (bottom left).
Furthermore, the dynamic range of each metric varies significantly, as the PIT/CDF, and KS are limited to [0,1], and the CvM, KLD, RMSE, and CDELoss have far more freedom.}
We also note that RMSE and Wasserstein distance have dimensions of redshift, while others are dimensionless, which makes the level of significance further incomparable between the metrics.}
This \cwr{figure} illustrates well the need for several different metrics to assess performance\revone{}{, and shows that any attempt to construct a ``super-metric" from a combination of these metrics should be done with extreme care if at all}. 



\section{Results \& Discussion} \label{sec:results}


Below we examine in detail the \revone{results}{values} obtained for \revone{each estimator using}{} each metric \revone{}{as evaluated over each estimator's $p(z)$ results for many cases of training set imperfection}. 
\revone{We divide our analysis into two sections, considering inverse redshift incompleteness and spectroscopic degraders separately.}{}
We discuss which metrics expose (or do not expose) particular trends in performance, and which are systematically biased (or unbiased) by distribution shape. 
We ultimately seek to identify which \refresponse{set of} metrics provide the most complete picture of photo-$z$ algorithm performance in our examined cases, rather than to identify (or tune) a ``best'' algorithm. 
Our list of photo-$z$ algorithms is not exhaustive, nor do we manipulate any algorithmic tuning parameters, thus we wish to emphasize that our work serves as an exploration of how one should evaluate estimators under various data challenge conditions. 
\revone{}{We divide our analysis into two sections, considering inverse redshift incompleteness and spectroscopic degraders separately.}







\subsection{Inverse Redshift Incompleteness Effects}

We begin by examining the effects of the inverse redshift incompleteness degradation stage, as a toy example of spectroscopic degradation. As the effects of this stage have been examined before, we perform this analysis largely with the goal of reproducing conclusions drawn in previous work. In \citet{Stylianou22}, the authors analyze the performance of \texttt{GPz} under inverse redshift incompleteness degradation. We reproduce this analysis, as well as examining the performance of our other estimators under this form of degradation. 


\begin{figure*}
  \centering
  \includegraphics[width=1.0\textwidth]{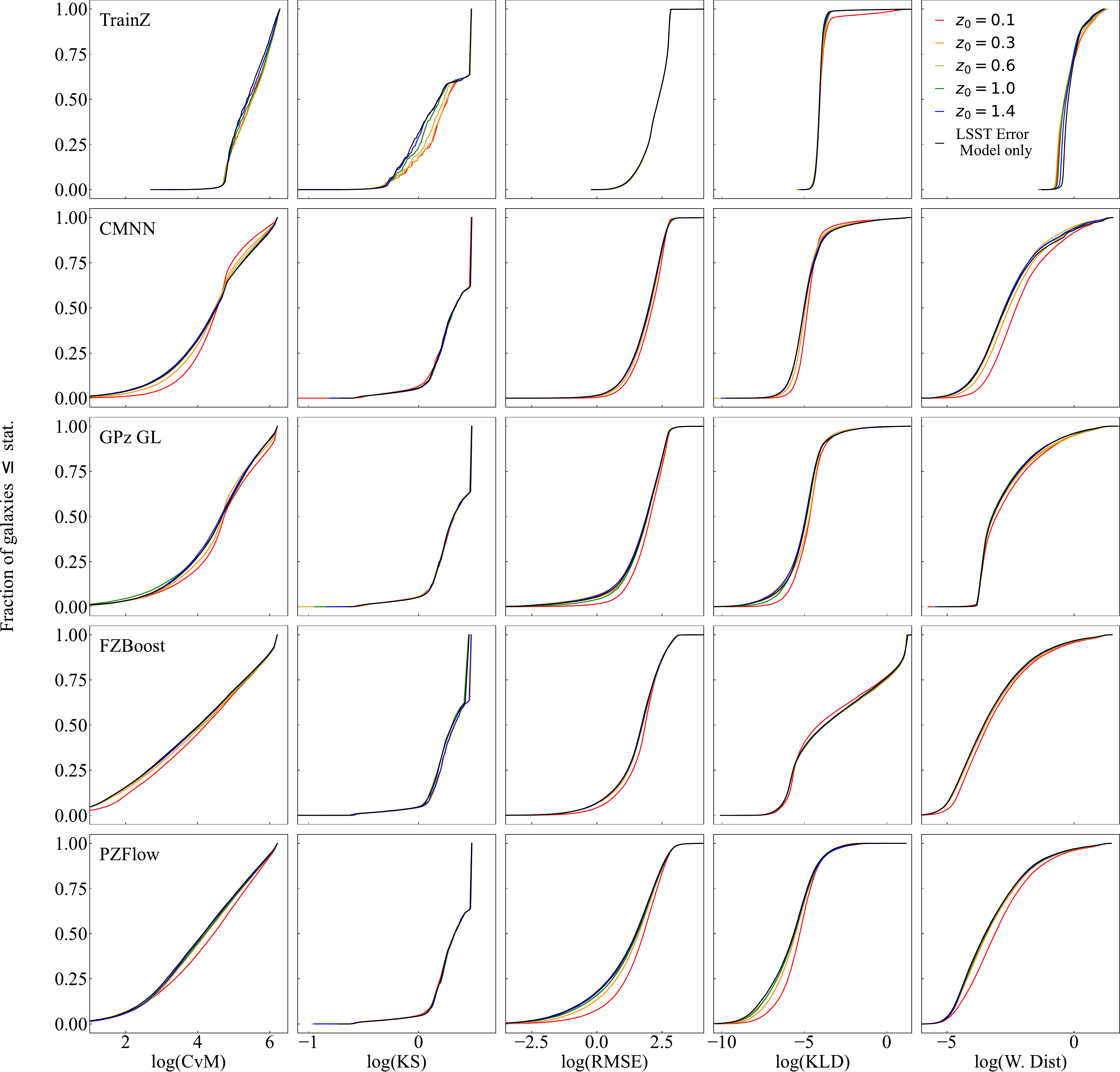}
  \caption{Quantile plots of the distributions of logarithmic CvM, KS, RMSE, KLD, and Wasserstein Distance statistics (columns) for each photo-$z$ estimation algorithm (row), under various degrees of inverse redshift incompleteness severity (different colored curves in each panel). 
  The $y$ value for each point on a curve is the fraction of outputs with a statistic value less than or equal to its corresponding $x$ value. 
  Rapid increases in slope correspond to a large number of galaxies having the same statistic value. 
  We expect cases with more severe degradation to have higher overall statistic values, and thus steeper slopes towards the right side of their axes. We note that these metrics expose very little difference in performance as $z_0$ is varied, with the exception of the most severe cases of $z_0=0.1$.
  }
  \label{fig:invz d2d} 
\end{figure*}

\subsubsection{CvM, KS, RMSE, KLD, W.\ Distance}

We find that the effects of inverse redshift incompleteness degradation on the majority of distribution-to-distribution statistics are minimal, across all photo-$z$ estimators. 
Our distribution-to-distribution statistic results are summarized in Figure~\ref{fig:invz d2d}, which shows the CDF of the distribution of per-galaxy values for each metric and photo-$z$ estimator. 
Comparing across \revone{panels in a fixed}{each} row \revone{provides a comparison between the results of different statistical tests}{shows the behavior of each metric} for a \revone{particular}{given} algorithm\revone{}{, whereas comparing down each column shows how a single metric differs between algorithms}. 
\revone{}{At first glance, Figure~\ref{fig:invz d2d} shows that for most metrics, there is greater sensitivity to photometric redshift algorithm than to the degree of inverse redshift completeness degradation of the training set.}






Examining row 1, we see how the different statistics \revone{examined}{considered} here characterize \texttt{TrainZ} performance. Little variation across different values of $z_0$ is seen, with only the most severe cases (low $z_0$) causing notable deviation in the \revone{CvM}{KS test}. Comparing the top row against other rows, we see that as we expect, statistic values for \texttt{TrainZ} are higher overall than those of our non-control estimators in the \revone{KS}{CvM}, RMSE, KLD, and W.\ Distance, but surprisingly slightly lower in the KS test. \texttt{TrainZ}'s apparent `good' performance in this statistic is a prime example of why using multiple performance metrics is important in assessing photo-$z$ estimator performance. 


\revone{}{Across our other estimation algorithms, shown in rows 2-5, these metrics illustrate remarkably little difference in performance for different $z_0$ values, with only the most extreme cases of degradation being distinguished.  The KS test appears to hardly differentiate at all between any levels of inverse redshift incompleteness. We note the significant spike at high KS values, indicating a pileup of cases in which this statistic indicates poor performance. This spike likely contains a combination of true catastrophic outliers and distributions that are narrow and slightly offset from the truth and thus yield artificially high KS test values. 
As both situations result in output distributions with little to no overlap with the truth, they both yield near-maximum KS values.}

\begin{figure*}
  \centering
  \includegraphics[width=0.7\textwidth]{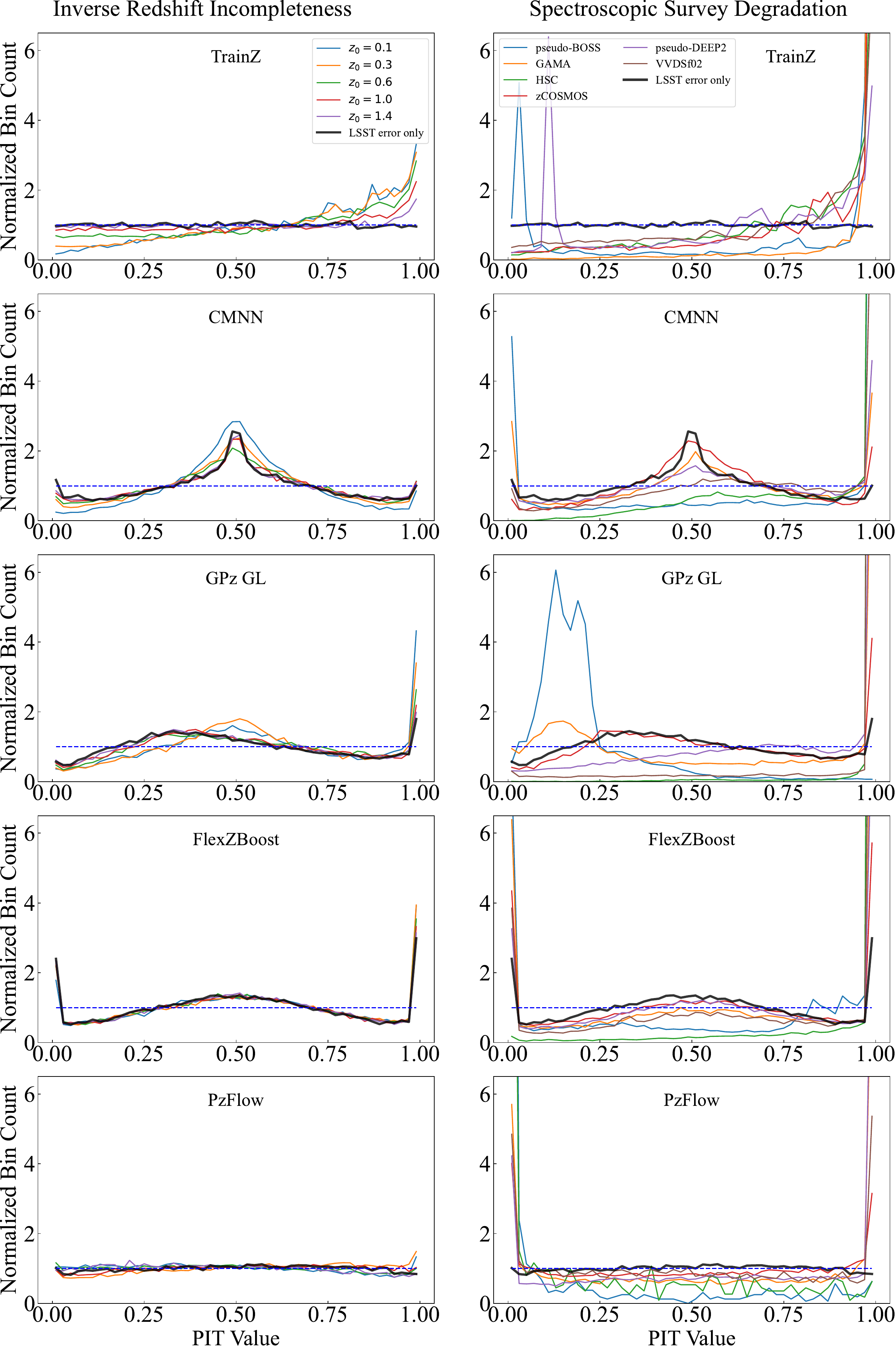}
  \caption{Probability integral transform distributions for inverse redshift incompleteness (left) and spectroscopic survey degradation (right). 
  A uniform distribution (ideal) is shown as a dashed line. We observe that each photo-$z$ estimator (shown in separate rows) has a distinctive and identifiable PIT shape. Shifts in this shape as degradation conditions are varied are more noticeable for Gaussian estimators than for \texttt{FlexZBoost} or \texttt{PzFlow}, though all experience a marked increase in side spike height under severe degradation, as expected. 
  }
  \label{fig:pit} 
\end{figure*}

\subsubsection{PIT}

In \citeDCt, the PIT and metrics thereof provided the most striking demonstration of the necessity of experimental controls like \texttt{TrainZ}, which consistently outperformed other estimation algorithms by the PIT under ideal conditions. As seen in the top left panel of Figure~\ref{fig:pit}, under biased training conditions \texttt{TrainZ}'s PIT curve is markedly shifted towards higher values. Thus in the presence of incomplete training data, \texttt{TrainZ} is no longer as useful as a control designed to optimize the PIT statistic. 


At all levels of degradation, \texttt{CMNN} (row 2) exhibits a clear middle bump, indicating that a disproportionate amount of its output $p(z)$ distributions are wider than the truth $p(z)$. 
The bump becomes more pronounced as degradation becomes more severe, without shifting to the left or the right. This again indicates that the primary effect of increasing the inverse redshift degradation on \texttt{CMNN} outputs is increased standard deviation. 

The \texttt{GPz} PIT distributions in the left column exhibit both a middle bump and a right side spike, indicating a variety of distribution shapes in its outputs. The right spike implies a large fraction of underestimated distributions, while the middle bump again indicates an excess of distributions that are wider than the truth $p(z)$. As $z_0$ is increased and degradation becomes less severe, the right side spike decreases. This makes sense: since higher redshift data are becoming more complete as $z_0$ increases, the algorithm is less likely to underestimate the redshift. However, it is interesting to note the lack of any left spike; this indicates that under inverse redshift incompleteness, \texttt{GPz} produces virtually no catastrophically high outliers compared to low ones. Decreasing $z_0$ also shifts the middle bump downwards from $0.5$, indicating that broad, slightly under-estimated distributions may be common under idealized training conditions. We note also that for $z_0 \gtrsim 0.6$, little further shift in distributions is observed, again indicating that increasing $z_0$ beyond this approximate value yields diminishing returns on photo-$z$ estimation improvement.

\texttt{FlexZBoost} and \texttt{PzFlow} exhibit  remarkably robust PIT behavior for a variety of levels of inverse redshift incompleteness; decreasing $z_0$ has hardly any affect on the PIT distribution shape for either estimator. \refresponse{This example illustrates that the PIT's utility as a diagnostic for redshift incompleteness depends on the photometric \refresponsetwo{redshift} algorithm, and the PIT needs to be used alongside other, complementary metrics.} \texttt{FlexZBoost} exhibits pronounced spikes on both sides, as well as a modest middle bump. The presence of spikes on both sides indicates that narrow, slightly offset distributions are common for \texttt{FlexZBoost}. The middle bump implies that diffuse distributions, or potentially distributions that are bimodal about the truth, are also prevalent. \texttt{PzFlow}'s PIT distribution is remarkably featureless, closely approaching a uniform distribution in all cases, with only minor deviations as degradation worsens. 

As \texttt{GPz} and \texttt{CMNN} are both restricted to Gaussian output distributions, the effects of degraded data on their outputs will be qualitatively quite different than in the cases of \texttt{FlexZBoost} and \texttt{PzFlow}. In \texttt{GPz} and \texttt{CMNN} outputs, there are only two tuneable output distribution parameters: mean and standard deviation. The PIT exposes how these are influenced effectively: lack of horizontal shifts in the \texttt{CMNN} PIT distribution, yet continuous increase in middle bump height as degradation worsens, indicates a trend in widening standard deviations with little shift in means, whereas with \texttt{GPz}, change in side spike size as well as shifts in the overall distribution likely indicate a combination of the two. 


\begin{figure*}
  \centering
  \includegraphics[width=0.95\textwidth]{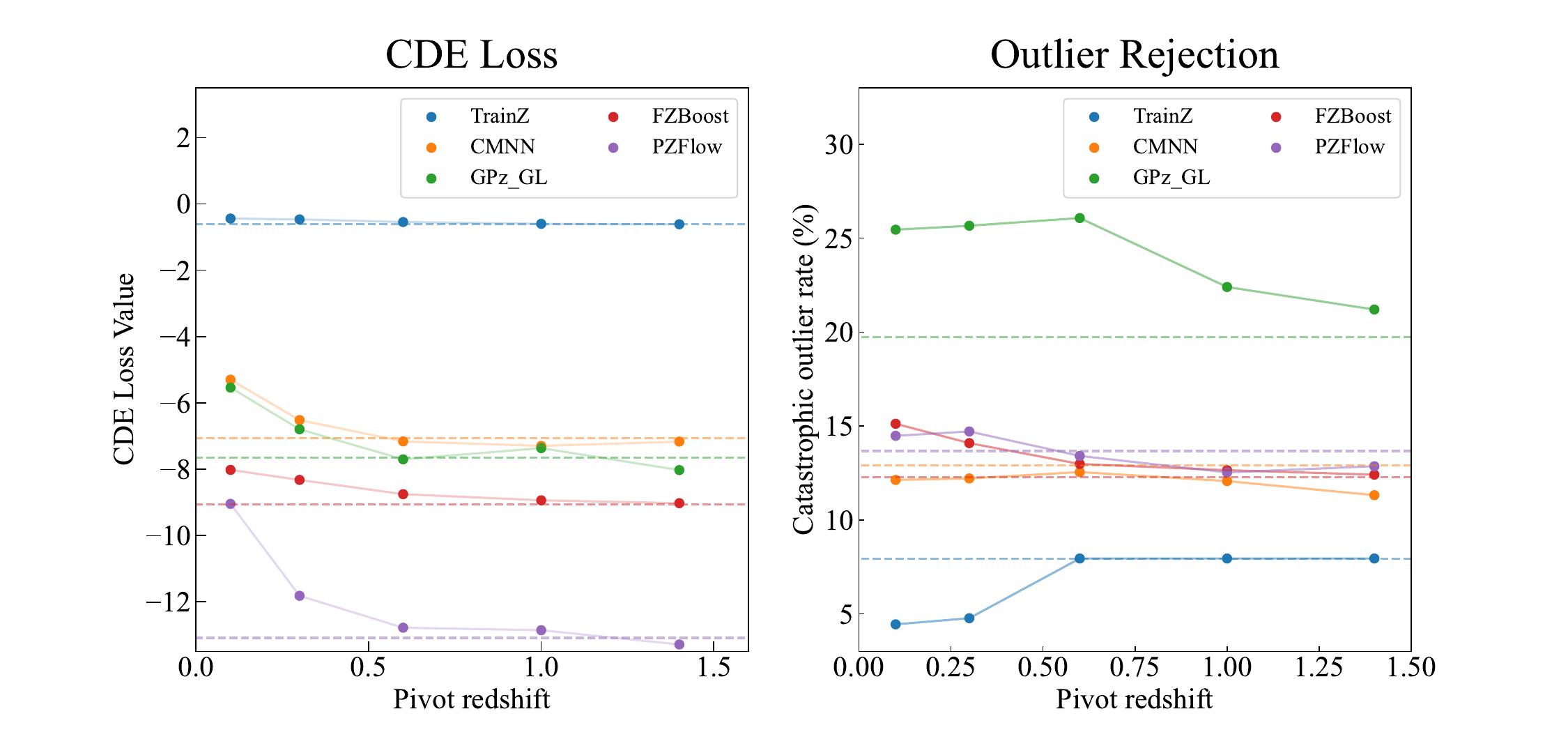}
  \caption{Plots of CDE loss values (left) and catastrophic outlier rates (right) as functions of the inverse redshift incompleteness pivot parameter, shown as solid lines. 
  Values for each estimator calculated under our control conditions (no inverse redshift incompleteness) are shown as dashed lines. We note that the CDE loss typically decreases as $z_0$ is increased, though we observe that the metric is far more sensitive to the estimator used than to the training set degradation. The catastrophic outlier rate decreases with increasing $z_0$ in some cases, but we see that this statistic is easily biased by scatter in point estimates, as the control estimator \texttt{TrainZ} appears to have done well.}
  \label{fig:invz CDE_OR} 
\end{figure*}

\subsubsection{CDE Loss}

The left panel of Figure~\ref{fig:invz CDE_OR} illustrates how the severity of inverse redshift incompleteness degradation affects the CDE loss metric. From examining the blue curve, we see that \texttt{TrainZ} has very weak evolution in the CDE loss as $z_0$ changes. Fortunately, the \texttt{TrainZ} CDE loss values are much higher (i.e., worse) than those for our non-control estimators, indicating that this statistic reflects galaxy-by-galaxy accuracy reasonably well. 

Results for \texttt{CMNN} and \texttt{GPz} (orange and green curves) are very similar to each other, with notable improvements as $z_0$ is increased, up to $z_0 \sim 0.6$. Though the trend is less pronounced, \texttt{FlexZBoost} exhibits mild improvements as $z_0$ increases. For lower values of $z_0$, \texttt{PzFlow} improvements are quite significant for increasing $z_0$. Though all non-control estimators exhibit some level of improvement in this statistic as degradation severity is decreased, results appear largely unaffected as $z_0$ is increased beyond $\sim$0.6 for all estimators. This suggests that increasing a theoretical pivot redshift beyond $z_0 \sim 0.6$ will result in minimal further improvements in the aspects of photo-$z$ estimation to which the CDE loss is sensitive. We also note that while the CDE loss is marginally sensitive to this form of training set degradation, it appears to be far more sensitive to the choice of estimator than to training set properties. 

\subsubsection{Outlier \revone{Rejection}{Identification}}

The right panel of Figure~\ref{fig:invz CDE_OR} shows trends in the catastrophic outlier rate as $z_0$ is increased for each of our photo-$z$ estimators. As discussed in Section~\ref{sec:outlier rejection}, this statistic is artificially low and not particularly meaningful for \texttt{TrainZ}. Our non-control algorithms provide more interesting results. \texttt{CMNN} and \texttt{PzFlow} show only weak trends here, as outlier rates fluctuate with respect to the baseline determined in the representative control case. For \texttt{FlexZBoost}, though the trend is slight, there is a clear decrease in outlier rate as $z_0$ is increased. 

As with its PIT and CDE loss, \texttt{GPz} seems to exhibit two distinctively different behaviors for $z_0\lesssim 0.6$ and $z_0 \gtrsim 0.6$. At low $z_0$, outlier rates are high, and approximately constant with $z_0$. However, there appears to be a decreasing trend in outlier rate as $z_0$ is increased past $0.6$. Unlike for the CDE loss, this statistic indicates that for \texttt{GPz}, there is continued improvement in performance as $z_0$ continues to increase past $0.6$. 


\subsubsection{Summary of metric results for inverse redshift incompleteness}

Overall, the effects of inverse redshift incompleteness degradation are small in comparison with the effects of spectroscopic survey degradation across all algorithms (see Section~\ref{sec:spec}). 
However, some metrics illustrate these small differences more effectively than others, and different photo-$z$ algorithms have differing sensitivity modes. The KS test, RMSE, and KLD reflect little in terms of the difference in performance across different $z_0$ cases, yet provide a helpful picture of overall estimation accuracy. The CvM test exposes small differences in performance across $z_0$ cases in the Gaussian photo-$z$ estimators, yet shows very little change for others. The PIT provides valuable insights into shifts in distribution shapes, particularly for Gaussian estimators, and illustrates two different behaviors for \texttt{GPz} in cases where $z_0 \lesssim 0.6$ and $z_0 \gtrsim 0.6$. The CDE loss illustrates clear performance improvement for all estimators up to $z_0 \sim 0.6$, yet shows little further improvement as $z_0$ continues to increase. Outlier rejection results are scattered for \texttt{CMNN} and \texttt{PzFlow}, while showing clear improvement for \texttt{FlexZBoost} as $z_0$ increases, and also capturing \texttt{GPz}'s differing behavior for $z_0 < 0.6$ and $z_0 > 0.6$


\subsection{Spectroscopic Survey Degradation Effects}
\label{sec:spec}

As anticipated, we find that the more varied and severe forms of incompleteness introduced with spectroscopic survey degradation have more significant effects on photo-$z$ estimation results than the toy form of degradation created by the inverse redshift incompleteness degrader.  
We naturally expect that more severe degradation should correspond to worse photo-$z$ performance that would be reflected in our metrics. As pseudo-BOSS is catastrophically incomplete, we expect it to fail more catastrophically than all other training sets, and produce both inaccurate and imprecise results. We  expect similarly poor performance from GAMA, though perhaps to a lesser degree as it is larger, and potentially biased performance under zCOSMOS and pseudo-DEEP2 degradation, placing precise estimates preferentially in the regions of redshift-magnitude space they cover with low accuracy for uncovered regions. 
As coverage is high under VVDSf02 and HSC degradation, we expect performance under these conditions to be both more accurate and more precise. 
We examine below the correlation between algorithmic performance across our various metrics and training set coverage of redshift-magnitude space and/or training set size.

\begin{figure*}
  \centering
  \includegraphics[width=1.0\textwidth]{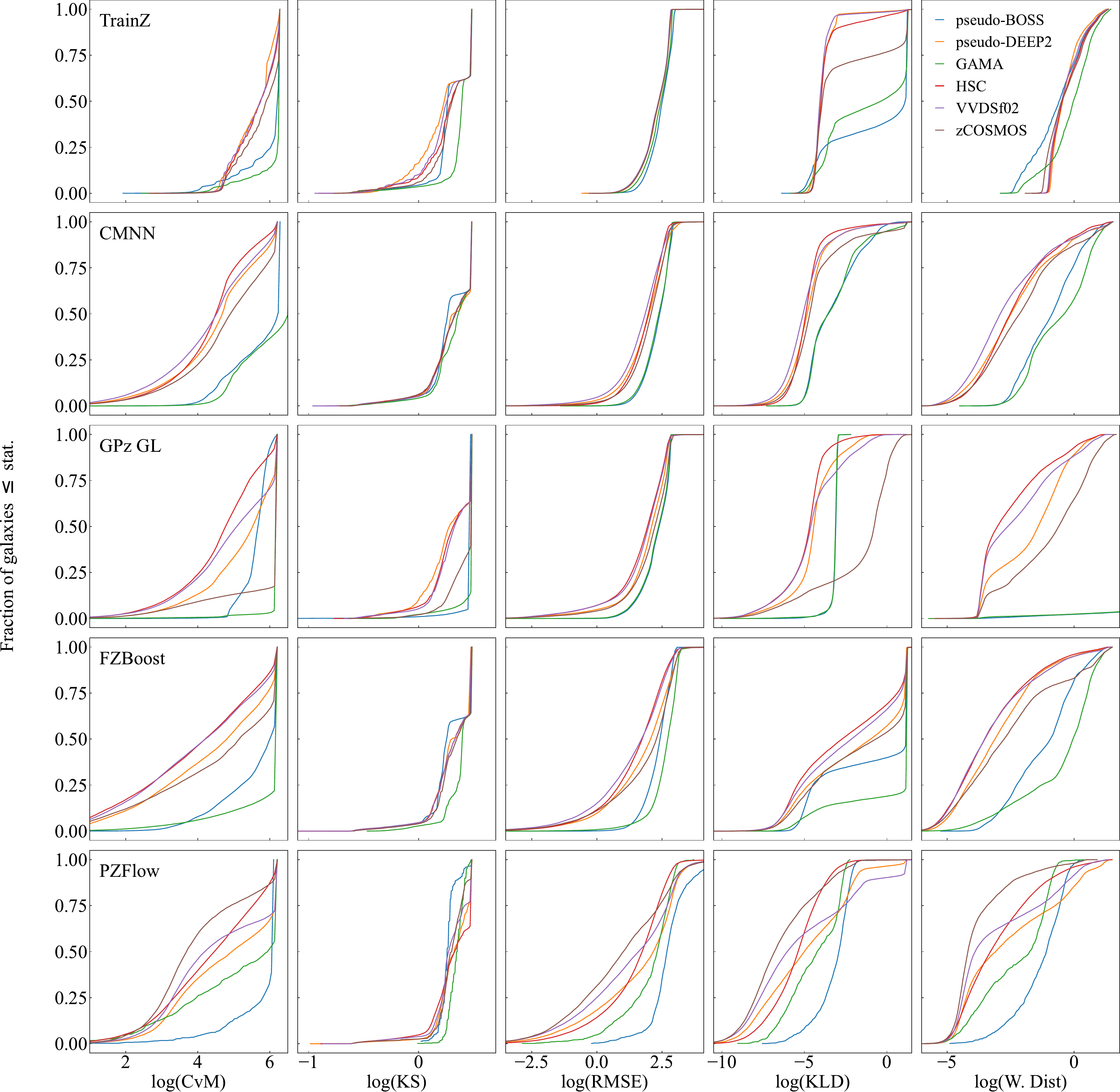}
  \caption{Quantile plots of the \refresponse{logarithmic} CvM, KS, RMSE,  KLD, and Wasserstein Distance statistics (columns) for each photo-$z$ estimation algorithm (rows), under each case of spectroscopic survey degradation (colors indicated in legend). Lines \refresponse{curving to the bottom right} indicate a larger proportion of high statistic values, and thus poorer performance. \refresponse{Galaxies with invalid \texttt{PzFlow} estimates are discarded to normalize the curves to unity.}  We see significant differentiation between the curves for \refresponse{different degraders for all metrics except for the KS test, and the performance patterns under degradation are mostly consistent across different photo-$z$ algorithms. We note that the BOSS and GAMA degraders result in particularly bad photo-$z$s, which is consistent with our expectation since these degraders only select bright training galaxies with a particular color cut. }
  }
  \label{fig:spec d2d} 
\end{figure*}

\subsubsection{CvM, KS, RMSE, KLD, W.\ Distance}


Figure~\ref{fig:spec d2d} summarizes the distribution-to-distribution statistics for each spectroscopic degrader and photo-$z$ estimator. 
In contrast with the inverse redshift incompleteness degradation, we see significant variation between the different degraders in practically every case, and the metrics have differing sensitivity to the different degraders' impact on each estimator. 

\texttt{TrainZ} performance (shown in row 1) seems to be strongly correlated with the number of galaxies in its training set (provided in Figure~\ref{fig:spec CDE plot} \cwr{and Table}~\ref{tab:spec rej}) and affected minimally by training set coverage, as indicated by \refresponse{similar} performance in the pseudo-DEEP2 and VVDSf02 cases \refresponse{as for} the HSC case in several metrics. 
The CvM and KLD show this particularly clearly. 
The RMSE provides no means of differentiating between different forms of incompleteness for \texttt{TrainZ}, while the KS test and Wasserstein Distance show little variation except in the cases of pseudo-BOSS and \refresponse{GAMA}, where large fractions of redshift-magnitude space are missing from the training data.

\texttt{CMNN} results are provided in row 2. 
We see that all  metrics \refresponse{except for the KS test} can discriminate between the two catastrophically incomplete training sets (with the pseudo-BOSS and GAMA degraders) and the other four. 
Low CvM values for these two cases are likely explained by wide output distributions, i.e.~the effect illustrated in Figure~\ref{fig:stdev fx}. 
There is \refresponse{less} variation in photo-$z$ performance when applying the pseudo-DEEP2, HSC, VVDSf02, and zCOSMOS degraders as quantified with any of the other metrics, though \refresponse{all but the KS test reveal some differences between the degraders}. 
The relatively minor effect on performance in the zCOSMOS case indicates that \texttt{CMNN} is successful at interpolating to regions of redshift-magnitude space that are only minimally covered by training data. 

\texttt{GPz} results are provided in row 3. 
The distributions of metric values shift significantly across our training sets, though the RMSE is \refresponse{the least sensitive of the metrics shown}. 
The other statistics, however, capture well the surprisingly poor performance of \texttt{GPz} on the zCOSMOS training set. 
Notably, the Wasserstein Distance captures a clear correlation with average training set coverage, as seen by examining the left-to-right ordering of the curves. 
As expected, estimation was largely unsuccessful on the pseudo-BOSS and GAMA cases, as shown by the significant deviation of the blue and green curves from the rest, which is particularly obvious in the KLD and Wasserstein distance.
It may be surprising how little the two cases differ from each other, but in fact the vast majority of estimates for both cases are what we may consider to be `failures' (see Section~\ref{sec:failure rates}): 
when a test galaxy is inadequately represented in training data, \texttt{GPz} returns so large a variance that the distribution is broad enough to appear to be a nearly flat line at $p(z) = 0$ for all $z$.
Thus the localized spike for pseudo-BOSS and GAMA distributions present across most of our statistics corresponds to the value each takes on in the case of a ``flat'' distribution compared with a narrow, unimodal distribution. 
The fact that zCOSMOS has under-performed or performed similarly when compared with these cases indicates that a significant fraction of its outputs are truly incorrect estimates not flagged by the algorithm as failures. 
This demonstrates the need for multiple metrics, specifically the interpretation of outlier/failure rates in addition to the distribution-to-distribution metrics.

\texttt{FlexZBoost} results are presented in row 4. 
Again results are split between pseudo-BOSS and GAMA versus the other four surveys, though the two catastrophically incomplete cases differ significantly from each other in several metrics. 
Perhaps surprisingly, GAMA estimation appears less successful than pseudo-BOSS estimation according to most metrics. GAMA curves (green) are shifted notably to the right in the \refresponse{CvM}, KS test, KLD, and Wasserstein distance when compared with pseudo-BOSS curves, indicating a larger proportion of high statistic values.
Though \texttt{FlexZBoost} does not have a well-defined notion of failure, distributions corresponding to galaxies that are not well-represented in training data are often very broad or multi-modal, appropriately indicating uncertainty. 
The vast majority of pseudo-BOSS outputs are such distributions, whereas a larger proportion of GAMA outputs are likely truly inaccurate estimations, similar to the zCOSMOS case with \texttt{GPz}. 
\texttt{FlexZBoost} estimates with the zCOSMOS degrader appear mildly worse than those for HSC, VVDSf02, or pseudo-DEEP2 degraders, particularly in the Wasserstein Distance, though the difference is not extreme.  

Row 5 presents \texttt{PzFlow} results. \refresponse{As noted in the caption, galaxies with invalid \texttt{PzFlow} estimates are discarded so the curves normalize to unity.}
It is worth noting a distinction between the results for pseudo-BOSS and GAMA: 
\texttt{PzFlow} produced similar numbers of estimates for each case, but with much higher accuracy for the GAMA training set. 
The increase from the $\sim$$10\%$ average coverage in pseudo-BOSS to the $\sim$$25\%$ \refresponse{coverage of GAMA 
significantly improved the accuracy}. This is clearly visible in the leftward shifts of the GAMA curve in the CvM and RMSE results shown in Row 5 of Figure \ref{fig:spec d2d}, but is made the most obvious by examining the CDE loss (see Figure \ref{fig:spec CDE plot}). 

Across all estimators, the \refresponse{CvM}, KLD and Wasserstein Distance consistently provided an informative \refresponse{view on how algorithm performance depends on the training set incompleteness}, while the RMSE \refresponse{and KS test show little (or inconsistent)} variation across different degrees of training set degradation, except in catastrophically poor cases that are unlikely to occur in realistic data. 
Due to its sensitivity to distribution tails, the CvM test at times distinguishes differences in performance not captured by other statistics, but as it is easily biased, 
it does not often provide an accurate picture of photo-$z$ estimator success on its own. 
For a few of the examined estimators, the KS test is \refresponse{relatively insensitive to training set incompleteness except in truly catastrophic cases, but as noted previously, it has the advantage of being} sensitive to shifts in distribution shape in the opposite way from the CvM and KLD tests (see Figure~\ref{fig:stdev fx}). 


\subsubsection{PIT}

Comparing the left and right columns of Figure~\ref{fig:pit}, we see that overall distribution shapes and trends in the PIT curves for the spectroscopic degradation cases are similar to those in the inverse redshift incompleteness cases, with the exception of the catastrophically incomplete pseudo-BOSS and GAMA training sets, and in the case of \texttt{GPz}, zCOSMOS as well. Side spikes become more prominent in the presence of severe degradation for all estimators (note the logarithmic scaling of the vertical axis in the right column of Figure~\ref{fig:pit}), though the PIT helpfully captures whether algorithms preferentially \cwr{produce outliers to low versus high redshift} when we compare the relative sizes of the left and right size spikes. For example, similarly-sized left and right side spikes for \texttt{CMNN}, \texttt{FlexZBoost}, and \texttt{PzFlow} indicate that catastrophically high and low outliers are roughly equally common from these alorithms, whereas the lack of a left side spike for \texttt{TrainZ} and \texttt{GPz} indicates that these algorithms are prone to under-estimation under these degradation conditions. 
This is an example of the clear illustration of trends in the specific way(s) in which algorithms can be inaccurate that the PIT provides, which is not provided by other metrics. 

\subsubsection{CDE Loss}

\begin{figure*}
  \centering
  \includegraphics[width=0.9\textwidth]{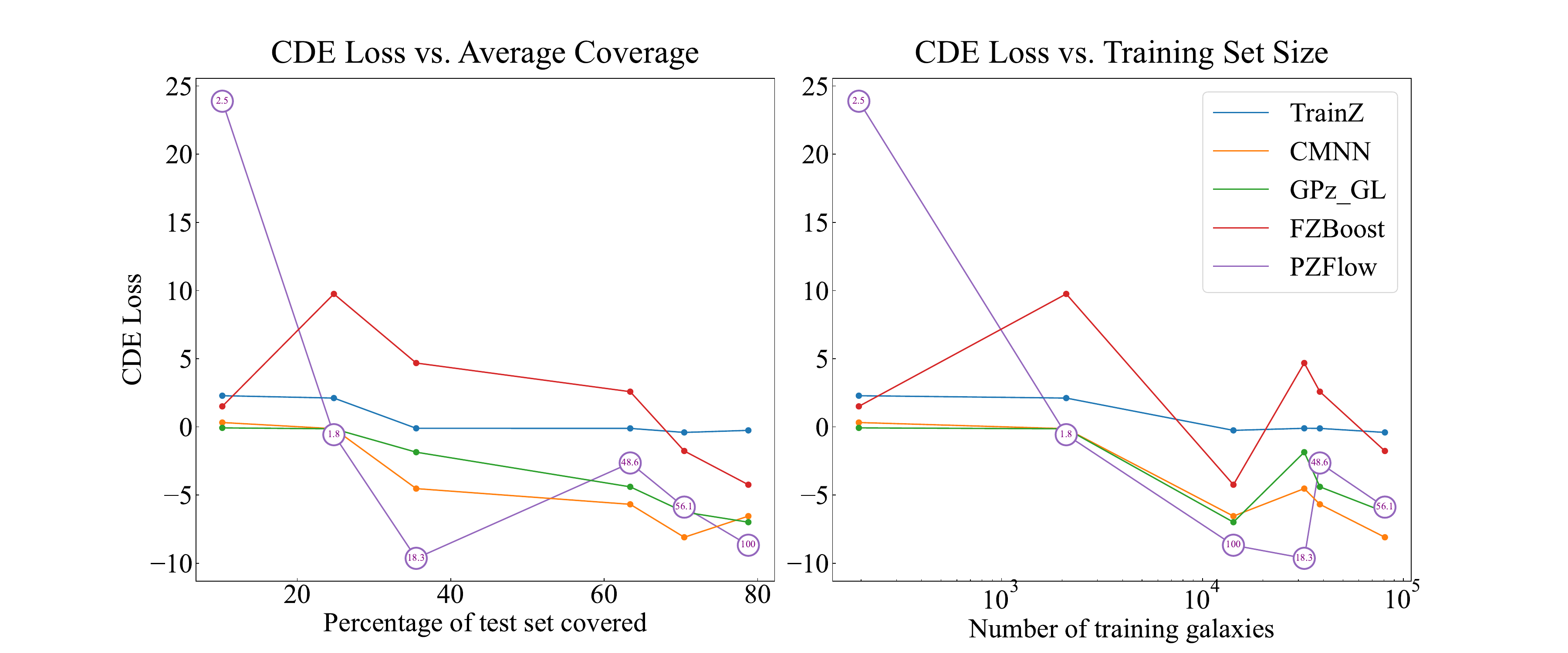}
  \caption{CDE loss plotted both as a function of training set coverage of the test set (left) and training set size (right). For most estimators, we see a relatively smooth downward trend in CDE loss as coverage increases, whereas we see much more scatter when considering training set size alone. 
  Since \texttt{PzFlow} has failed to produce an estimate for large fractions of most of our training sets, this \texttt{PzFlow} data only accounts for the galaxies on which estimation successfully occurred. We have included the \texttt{PzFlow} success rate for each training set in the center of each \texttt{PzFlow} data point. }

\label{fig:spec CDE plot} 
\end{figure*}

Examining Figure~\ref{fig:spec CDE plot}, we see that many results shown in Figure~\ref{fig:spec d2d} are corroborated. 
Examining the left panel of Figure~\ref{fig:spec CDE plot}, we see a clear anti-correlation between training set coverage and CDE loss value for \texttt{CMNN} and \texttt{GPz}. \texttt{TrainZ} shows very weak evolution as training set coverage increases past around $25\%$. With the surprising exception of the pseudo-BOSS case, \texttt{FlexZBoost} also shows a clear decrease in CDE loss as coverage is increased. The trend in \texttt{PzFlow} results is not as clear, though note that only galaxies for which successful estimation was performed contribute to this statistic. Nevertheless, with the exception of the GAMA case, a monotonically decreasing trend in CDE loss appears as coverage is increased. The surprisingly low zCOSMOS value is likely due to high accuracy in the smaller number of estimates made, but does not reflect the failure rate of over $80\%$. 

Examining both coverage and training set size, performance is obviously poor in our catastrophically incomplete cases. Though there is a downward trend as both coverage and training set size are increased, the coverage trend is much clearer, showing monotonic or nearly monotonic decreases in all but the \texttt{PzFlow} case. Though there is also a decreasing trend in CDE loss as we increase training set size, the scatter is much more significant. 
All non-control algorithms show the most significant scatter among the larger training sets, indicating that training set size alone is not an necessarily an accurate indicator of potential performance success without also considering coverage of redshift-magnitude space.

Finally we also note that, similar to the inverse redshift incompleteness case, in many instances differences between estimators are exposed more starkly by this metric than differences across training sets. While this metric may be successful at distinguishing estimators from one another, it is not as indicative of the effects of training set imperfections as some others examined here. 

\subsubsection{Outlier Rejection}

\begin{table*}
\begin{center}
\begin{tabular}{||c c c c c c||} 
 \hline
 Degradation Type: & \texttt{TrainZ} & \texttt{CMNN} & \texttt{GPz} GL & \texttt{FlexZBoost} & \texttt{PzFlow}\\ [0.5ex] 
 \hline\hline
 pseudo-BOSS (195 galaxies): & 7.133 & 10.19 & 25.45 & 15.97  & 2.644 $\&$ 97.45 $\rightarrow$ 97.51\\
 \hline
 GAMA (2,099 galaxies): & 4.410 & 3.778 & 26.07 &  3.049 & 0.535 $\&$ 98.21 $\rightarrow$ 98.22\\
 \hline
 HSC (14,253 galaxies): & 4.432 & 16.03 & 26.25 & 17.60 & 19.78 $\&$ 0.0 $\rightarrow$ 19.78\\  
 \hline
 zCOSMOS (32,089 galaxies): & 5.504 & 25.97 & 0.0 & 24.57 & 16.05 $\&$ 81.67 $\rightarrow$ 84.61\\  
 \hline
 pseudo-DEEP2 (38,385 galaxies): & 7.027 & 15.43 & 25.66 & 15.68 & 41.78 $\&$ 51.38 $\rightarrow$ 71.61\\
 \hline
 VVDSf02 (80,845 galaxies): & 8.186 & 21.17 & 39.62 & 20.68 & 37.88 $\&$ 43.86 $\rightarrow$ 65.13\\
 \hline\hline
 LSST Error Model only (30,953 galaxies): & 7.949 & 12.91 & 19.75 & 12.27 & 13.67 $\&$ 0.0 $\rightarrow$ 13.67\\ [1ex] 
 \hline
\end{tabular}
\caption{Table showing percentage of estimates considered catastrophic outliers via outlier rejection for estimators trained on spectroscopic survey degraded data. For \texttt{PzFlow}, two numbers are shown: the first is the catastrophic outlier rate among successful estimates, the second is the percentage of the test data for which no estimate was made. The third number is the percentage of the test set that either produced a catastrophic outlier or a failed estimation. It is clear that outlier rate does not display a reliable correlation with the variation of training set perfection, either as size or coverage are changed. }
\label{tab:spec rej} 
\end{center}
\end{table*}

As discussed in Section~\ref{sec:outlier rejection}, the outlier rejection metric is easily biased low by large standard deviations in output point estimates. Examining \refresponse{Table}~\ref{tab:spec rej}, we do see this reflected in our results. The outlier rejection metric produces low values for several cases in which we would expect poor performance, due to very large scatter in the results.
\texttt{TrainZ} values differ little from those in the inverse redshift incompleteness case, as we expect. This effect also explains some surprisingly low pseudo-BOSS and GAMA results, and the `0.0' for \texttt{GPz}'s zCOSMOS case. 
However, we do not simply see an inverse correlation with outlier rate and performance either, as in some cases this metric appears high where we would expect it to be low, or does indeed perform as expected. 
Due to the multiple features of the distributions of outputs that affect this statistic, we do not find a simple relationship between outlier rate calculated via this method and either training set coverage or size. In cases where the distribution of mode estimates is supposed {\em a priori} to be a narrow cluster about the diagonal, this statistic provides an informative description of catastrophic outlier rate. However, in cases of poor estimation, this statistic is easily biased towards being unrealistically low, and therefore is not a reliable indicator of performance in such cases. 

\subsection{Failure Rates}
\label{sec:failure rates}

If an algorithm is tested on data that it judges to be far enough from its training set as to make its predictions unreliable, it would be preferable for it to fail to perform any estimation or return a distribution that adequately indicates this uncertainty, as opposed to taking a wild guess. 
Indicators of unreliability are in a sense a metric that an algorithm can deliver to users, and different algorithms do this to varying degrees.
Here we discuss some of the ways such flags manifest among the algorithms we tested and how effective we found those indicators to be.

As seen in the rightmost column of \cwr{Table}~\ref{tab:spec rej}, \texttt{PzFlow} rarely produces estimates for data that is not well-represented in its training set, with the exception of small gaps. 
This would make \texttt{PzFlow} an ideal choice in situations where accuracy is prioritized over completeness of photo-$z$ estimation.

For the other examined algorithms, quantifying a `failure rate' is not as straightforward, and would necessitate further work. With \texttt{GPz} for example, a certain mean or standard deviation could be used to establish a cutoff above (or below) which an estimate is considered a `failure', as \texttt{GPz} returns distributions with identifiably 
extreme parameters under poor training set representation. This leaves the judgment of success somewhat up to the user, though it may necessitate further post-estimation data processing. 

\texttt{FlexZBoost}'s notion of `failure' seems nebulous, though distribution multimodality may serve as a useful metric in cases of extremely poor training. Severely multimodal distributions could be chosen to constitute failures, but again this requires additional data examination after estimation is complete, and identifying multimodal distributions is not as easily implementable as a mean or standard deviation cutoff in the case of \texttt{GPz}. Given that even under GAMA training set conditions, such multimodal distributions were rare, \texttt{FlexZBoost} is ineffective at identifying when it cannot make an accurate estimate.

\texttt{CMNN} has little to no notion of estimation ``failures'' in its outputs: most distributions have passably reasonable parameters, so `failed' estimations cannot be identified as those with means of $\sim$$200$ or standard deviations of $\sim$$50$ for example, as in the case of \texttt{GPz}. This hinders accuracy under poor training set conditions, without making this obvious to the user. 

\texttt{TrainZ}, by definition, never reports a notion of unreliability.

\section{Conclusion}
\label{sec:conclusion}

\refresponse{Incompleteness and non-representativeness of spectroscopic training sets are common problems for photometric redshift estimation with deep imaging surveys.  In this work, we have investigated how a variety of metrics commonly used to quantify the performance of photometric redshift algorithms in RAIL react to (and can serve as diagnostics of) imperfections in spectroscopic training samples.  We considered the case of inverse redshift incompleteness, and non-representativeness associated with a variety of spectroscopic surveys.  The photo-$z$ performance metrics were applied to multiple different photo-$z$ algorithms -- without attempting to optimize or fine-tune their hyper-parameters -- as a demonstration of how these metrics react to training set imperfections.}

It is clear from our work that spectroscopic non-representativity effects have more severe consequences for \refresponse{photo-$z$} estimation accuracy than incompleteness effects, regardless of the algorithm being used. We conclude that, as expected, inverse redshift incompleteness degradation alone does not realistically approximate the complexity of real training data.
We also clearly demonstrate the need for multiple metrics in order to make a complete and accurate assessment of an algorithm's performance. {Here we recapitulate our conclusions for individual metrics:}

\begin{itemize}
\item The RMSE is affected comparatively little by the kinds of degradation we have examined. Only in the cases of catastrophically incomplete spectroscopic survey degradation is it significantly affected. It provides little information on algorithm performance in comparison with the other metrics examined here. As is examined in \citet{malz_approximating_2018}, the RMSE lacks sensitivity to distribution tails, and horizontal shifts of output distributions (which significantly affect tails) are very common under our degradation conditions. Thus the RMSE misses much of the changes in estimation behavior. 
\item The CvM test is quite sensitive to changes in training data, likely due to its high sensitivity to distribution tails. However, it is heavily biased by distribution shape, and often gives an inaccurate representation of performance. In several cases CvM values are (misleadingly) lower for instances of poor performance (such as when training with the pseudo-BOSS/GAMA surveys), and high where we know estimation is more accurate. It is an effective tool for picking up small variation in output distributions as training data are changed, but does not reflect estimator accuracy well on its own. 
\item Though the KS test is also susceptible to biases induced by distribution shape, we note that it may be informative when combined with other metrics, as it responds differently to systematic bias than either the CvM or KLD, as shown in Figure \ref{fig:stdev fx}. 
\item The KLD adequately \refresponse{reflects} differences in performance between spectroscopically degraded training sets. There is an obvious correlation between KLD value distributions and training set coverage, and we note that this statistic has also successfully identified some cases with high rates of inaccurate estimation. 
\item The Wasserstein Distance is also reliably representative of algorithm performance. It illustrates a clear correlation between training set coverage and performance for several algorithms, which is not shown as clearly by other metrics. 
\item The PIT reveals valuable information about distribution shapes not necessarily captured by other performance metrics, as it illustrates whether distributions are characteristically too wide (``middle bumps'') or too narrow (``side spikes''). 
\item The CDE loss metric shows modest correlation with training set coverage for most estimators, and provides a ``single number evaluation'' of performance. Though it does not necessarily provide a reliable picture of performance in instances of high failure rates, it provides a concise assessment of performance and is quick to implement. However, it seems to be more sensitive to differences between algorithms than training set imperfections. 
\item While catastrophic outlier rate is a very helpful metric with which to assess algorithm performance, implementing it via outlier rejection produces varied results. In cases of very poor estimation, outlier rejection produces an artificially low outlier rate, thus giving an inaccurate picture of performance. We recommend using a different method to identify catastrophic outliers in such cases. However, in cases where estimation produces a relatively tight distribution of point estimates, outlier rejection can be a valuable technique for identifying catastrophic outliers. 
\end{itemize}

We also note that further work on the quantification of failed estimation with \texttt{CMNN} and \texttt{FlexZBoost} is necessary in order to make their results optimally useful. 
This will be especially interesting in future work that considers estimation method optimization and different sample selections, which could be quite relevant to estimation failures.

\cwr{Future work could also employ these metrics in the context of statistical approaches to understand the impact of covariate shifts due to mismatches between training and testing sets, for example the stratified learning approach of \cite{https://doi.org/10.1002/sam.11643} and \cite{2025OJAp....8E..50M} that relies on analysis of subsets of the training and test sets prior to considering the full sample.}


In this work, we have shown the strengths of various statistical metrics to quantify the performance of photo-$z$ algorithms in the context of relatively simple test cases: a handful of photo-$z$ algorithms with default parameter settings, and with a limited illustrative set of training set imperfections.  The outcomes demonstrated here can serve as the basis for future work that will be important to carry out before analysis of LSST data using photo-$z$.  For example, a more thorough method comparison while allowing for estimation method parameter tuning will be an essential exercise.  In this work, we considered 10-year LSST depth, but studies prior to early LSST analyses will need to consider more shallow and less homogeneous datasets such as those that will be acquired in the first year of the survey. Here we considered a \refresponse{fixed} test set, but in reality it will be useful to explore how sample selection might usefully be varied to accommodate the performance of photo-$z$ estimation algorithms.  It will also be important to consider the impact of blending of galaxy light profiles, which requires a study based on simulated or real imaging rather than the extragalactic catalogs used in this work.  \cwr{Finally, we note that the models for the extragalactic galaxy sample in \citet{2025MNRAS.544.3799O}, as used for this work, have continued to evolve after publication of that work, and tests on a future version with updated matching of galaxy color-redshift relations would be valuable. } For all of these cases, however, the intuition developed in this work about how statistical metrics can be used to understand photo-$z$ performance will be an important starting point.

\section*{Acknowledgments}
AC led the analysis including most code implementation with guidance and contributions from co-authors, drafted initial paper version, participated in subsequent editing. AIM advised the lead author, contributed to conceptualization, methodology, project administration, resources, validation, and writing; DESC Builder for past contributions to RAIL codebase.  TZ implemented distribution metrics, advised RAIL running and analysis, edited the manuscript. RM advised on project progress and presentation, edited paper. OL contributed to RAIL and to project development and execution. FB assisted with data selection and pruning.  SJS contributed to RAIL code, assisted in some RAIL implementation details.  JFC contributed to RAIL, assisted with use of PZFlow.  IM wrote HSC degrader used to generate one of the training samples.  DO contributed to RAIL and to project development and execution. QH and JC-T contributed to RAIL.


This paper has undergone internal review in the LSST Dark Energy Science Collaboration. 
The internal reviewers were Chad Schafer and Chihway Chang. 
We thank Jeff Newman for helpful feedback on the details of the spectroscopic surveys we emulated and Tom Loredo for the suggestion to evaluate the Wasserstein metric. \cwr{We also thank Roberto Trotta for helpful feedback on the manuscript.}

AC was supported by an LSST-DA grant funding student travel to the 2023 Rubin Project and Community Workshop.
AIM, RM, TZ, OL, and DO are supported by Schmidt Sciences. 
RM is supported in part by the Department of Energy Cosmic Frontier program, grant DE-SC0010118. FB is supported for this work by NASA under JPL Contract Task 70-711320, ``Maximizing Science Exploitation of Simulated Cosmological Survey Data Across Surveys''. IM is supported by the U.S.\ Department of Energy, Office of Science, Office of High Energy Physics Cosmic Frontier Research program under Award Number DE-SC0010008.  QH is supported by STFC grant ST/W001721/1.

The DESC acknowledges ongoing support from the Institut National de 
Physique Nucl\'eaire et de Physique des Particules in France; the 
Science \& Technology Facilities Council in the United Kingdom; and the
Department of Energy and the LSST Discovery Alliance
in the United States.  DESC uses resources of the IN2P3 
Computing Center (CC-IN2P3--Lyon/Villeurbanne - France) funded by the 
Centre National de la Recherche Scientifique; the National Energy 
Research Scientific Computing Center, a DOE Office of Science User 
Facility supported by the Office of Science of the U.S.\ Department of
Energy under Contract No.\ DE-AC02-05CH11231; STFC DiRAC HPC Facilities, 
funded by UK BEIS National E-infrastructure capital grants; and the UK 
particle physics grid, supported by the GridPP Collaboration.  This 
work was performed in part under DOE Contract DE-AC02-76SF00515.

\bibliography{references}{}
\bibliographystyle{mnras}

\appendix

\section{Detailed calculation of metrics}\label{app:metrics}

\subsection{Distribution-to-Distribution Metrics}
\label{app:distdist}

\begin{description}

    \item[\textbf{Root Mean Square Error (RMSE):}]
    The Root Mean Square Error, or Root Mean Square Deviation, is the quadratic mean of a set of differences between true values and predicted values. 
    For a set \revone{$n$ of }{of $\{z_{k}\},\ k = 1, 2, \dots, n$ redshifts at which } true values \revone{$X_i(z)$}{$p(z | x_{i})$ have been evaluated} and a set of $n$ predicted values \revone{$x_i(z)$}{$p(z | x_{i}, \pi_{j})$ at the same $\{z_{k}\}$}, the RMSE \revone{}{of the $i$th galaxy under the $j$th estimator} is defined as 
   \begin{equation} 
   \mathrm{RMSE}_{i, j} = \sqrt{\frac{1}{n}\sum_{k=1}^{n}\left(p(z_{k} | x_{i}) - p(z_{k} | x_{i}, \pi_{j})\right)^{2}}
   . 
    \end{equation}
    \revone{In our case, we have computed this statistic galaxy-wise, with the $X_i(z)$ corresponding to the true posteriors for each galaxy evaluated on a grid of redshift values, the $x_i(z)$ corresponding to our estimator outputs for each galaxy evaluated on the same grid, and $n$ corresponding to the number of redshift grid points.}{}
    We implemented this with the mean squared error function in \texttt{sklearn.metrics.mean\_squared\_error}, comparing true PDFs to estimated PDFs. 

    \item[\textbf{Kullback-Leibler Divergence (KLD):}]
    The KLD is another statistical \revone{distance}{measure of discrepancy}, used to measure the \revone{difference}{divergence} between a reference (true) probability distribution and \revone{a second}{an approximating} (estimated) distribution\revone{}{, in the form of the information lost by using the approximation instead of the truth}. 
    \revone{Borrowing the notation from above, i}{I}t is defined point-wise for each \revone{$i$}{$k$} as 
    \begin{equation}
    \mathrm{KLD}_{i,j,k} = p(z_{k} | x_{i}) \ln\left[\frac{p(z_{k} | x_{i})}{p(z_{k} | x_{i}, \pi_{j})}\right]
    ,
    \end{equation}
    though precise functional forms may vary slightly in different implementations of the statistic. 
    \revone{}{It should be noted that, unlike the RMSE, the KLD is asymmetric, meaning the value would be different if we switched the positions of the true and approximating distributions;
    further detail may be found in the appendix of \citet{malz_approximating_2018}.}
    
    We employed the \revone{KL-Divergence}{KLD} function with \texttt{scipy.special.kl\_div} (\citealt{KLD}), comparing true PDFs evaluated on a discrete redshift grid to  estimated PDFs, evaluated on the same grid. 
    Since the \revone{KL divergence}{KLD} is only defined for nonzero values \revone{}{$p(\cdot)$} with \texttt{scipy.special}, all zeros resulting in evaluating the PDFs over the chosen grid were set to $10^{-300}$ . 
    For each galaxy, this provided us with a \revone{set $K$ of KLD values at each of our $n$ grid points.
    We then performed a continuum approximation over our $n$ grid points,}{grid of $n$ KLD values summed as $\mathrm{KLD}_{i,j} = \sum_{k=1}^{n}\mathrm{KLD}_{i,j,k}$ to report a final value.}
\subsection{CDF-based metrics}
     

\revone{}{In our descriptions of metrics based on the cumulative distribution function (CDF), the integral of a PDF, we denote the CDF of each galaxy $i$'s true and estimated posterior as $F_i(z)$ and $f_i(z)$ respectively.
When the CDF is estimated from samples, it is technically an empirical CDF (ECDF), which we denote as $\hat{F}(z)$ if it is derived from the true redshift posteriors and $\hat{f}_{j}(z)$ if derived from the $j$th estimator's redshift posteriors.
Both CDFs and ECDFs are obtained from the statistical ensembles output by RAIL stages using \texttt{qp} \citep{malz_approximating_2018}.}
    
    \item[\textbf{Kolmogorov-Smirnov \revone{}{(KS)} Test Statistic: \revone{(KS)}{}}] 
    The KS defines a distance between a \revone{cumulative distribution function}{CDF}, in our case \revone{the CDF}{that} of the true posterior generated for the test data, and an \revone{empirical distribution function}{ECDF}, in our case an estimator's output.
    \revone{In our case, both such distributions have been obtained from the statistical ensembles output by RAIL stages using \texttt{qp}.}{}
    \revone{For a cumulative distribution function $F_i(z)$ and an empirical distribution $f_{i,n}(z)$, t}{T}he KS is defined as 
    \begin{equation}
    \mathrm{KS}_{i,j} = \sup_{k} |f_{i,j}(z_{k}) - F_{i}(z_{k})|.
    \end{equation} 
    \revone{where for $n$ independent identically distributed observations,
    \begin{equation}
    f_{i,n}(z) = \frac{\mathrm{number\ of\ observations\ \leq \ }z}{n}. 
    \end{equation}
    Again our indices $\{ i \}$ refer to each galaxy and our indices $\{ k \}$ refer to our redshift grid points, over which we evaluate our empirical distribution functions $f_{i, j}(z_{k})$.}{
    }
    The KS can take on values between 0 and 1, with values closer to 0 indicating a more uniform PIT distribution (see Sec.~\ref{sec:pointdist}).  
    We calculated this statistic using \texttt{scipy.stats.kstest}, comparing the true CDF evaluated on a discrete grid to the estimated CDF, evaluated on the same grid.
    Therefore we expect more representative and complete training sets to yield lower KS test \revone{}{statistic} values. 

    \item[\textbf{Cramér-von Mises \revone{Criterion}{} (CvM) \revone{}{Criterion}:}]
    The Cramér-von Mises criterion is an alternative to the KS Test\revone{,}{ Statistic} in which the mean square distance between distribution functions is obtained according to
    \begin{equation}
    \mathrm{CvM}_{i,j} = 
    \int_{-\infty}^{+\infty} [f_{i,j}(z) - F_{i}(z)]^2 \mathrm{d}F_{i}(z) .
    \end{equation}
    \revone{using the same notation as above.}{}We calculated \revone{this statistic}{the CvM criterion} using \texttt{scipy.stats.cramervonmises} (\citealt{CvM}) to compare the true CDF evaluated on a discrete grid \revone{}{$\{z_{k}\}$} to the estimated CDF.
    We also expect higher-quality training data to yield lower CvM values. 

    \item[\textbf{Earth-Mover's Distance (First Order Wasserstein Distance):}]
    The Wasserstein Distance is a measure of the distance-weighted integrated probability density of one distribution with respect to another distribution.
    In one dimension, the Wasserstein Distance becomes the same as the Earth Mover's Distance (EMD). 
    This unique name is in fact very illustrative of what the metric seeks to convey: 
    imagining the two probability distributions being compared as piles of sand (or earth), the EMD is designed to quantify the minimum cost of building the \revone{smaller}{second} pile out of sand taken from the \revone{larger}{first}. 
    This ``cost'' is defined as the amount of sand moved, multiplied by the distance it is moved by\revone{}{, quantified by the EMD, which reduces to the L1 norm of the distance between the CDFs of the distributions}. 

    The \texttt{scipy.stats.wasserstein$\_$distance} (\citealt{wass_dist_scipy}) implementation of this metric was used to calculate the distance between the PDFs of each galaxy's true posterior and estimator output, evaluated over a discrete redshift grid. 
    The implementation in question evaluates the EMD \revone{}{recursively} as \revone{$\text{EMD} = \sum_{j}\text{EMD}_{j}$ for $\text{EMD}_{j} = x_{i}(z_{j}) + \text{EMD}_{j-1} - X_{i}(z_{j})$ and $\text{EMD}_{0} = 0$.}{$\text{EMD}_{i,j} = \sum_{k}\text{EMD}_{i,j,kk}$ where
    \begin{equation}
        \mathrm{EMD}_{i,j,k} = p(z_{k} | x_{i}, \pi_{j}) + |\mathrm{EMD}_{i,j,k-1}| - p(z_{k} | x_{i})
    \end{equation}
    for $\text{EMD}_{i,j,0} = 0$.
    }
    Formally, we should then multiply by a constant $\mathrm{d}z$.  
    However, we used a discretized implementation without that factor (which is the same for each galaxy\revone{}{, estimator, and training set}) and so this quantity is only defined up to a constant.  
    Since we are only using it as a comparative metric, this is an acceptable choice.
\end{description}

\subsection{Distribution-to-Point Metrics}
\label{app:pointdist}


\begin{description}
    \item [\textbf{Probability Integral Transform (PIT):}]
    \label{sec:PIT}

    \revone{For some random variable $X$ with a cumulative distribution function $F_{X}$, the}{The CDF $F_{Z}(z') = \int_{-\infty}^{z'} p_{Z}(z) \mathrm{d}z$ of a random variable $Z$ is itself a} random variable \revone{$Y := F_{X}(x)$ has a uniform distribution}{$Y:= F_{Z}(z')$}. 
    In our case, estimators output Bayesian posteriors of the form \revone{$p = p(z\ |\ \mathrm{photometry})$}{$p(z | x_{i}, \pi_{j})$}, where the redshift $z$ is our random variable, from which \revone{cumulative distribution functions $F_z$}{per-galaxy CDFs $F_{Z_{i}}(z'_{i})$} may be obtained. 
    If these posteriors are perfectly consistent with \revone{}{their corresponding population of} true redshift values, \revone{we may replace $z \rightarrow z_\text{true}$}{evaluation of each CDF at $z' = z_{\text{true}}$ yields values $y_{i}\sim Y_{i}$}, \revone{and}{where} the distribution of the random variable \revone{$Y := F_{z}(z_\text{true})$}{$Y$, known as the probability integral transform,} will be uniform. 
    \revone{Thus we define a CDF for a single galaxy's distribution as\revone{: $ \int_{-\infty}^{z_\text{true}} {p(z\ |\ \text{photometry})\ } \mathrm{d}z$}{$\int_{0}^{z_\text{true}} p(z | x_{i}, \pi_{j}) \mathrm{d}z$},
   and the distribution of such CDFs to be the PIT.}{} 


    \item [\textbf{Conditional Density Estimate (CDE) Loss:}]
The CDE Loss is an extension of the RMSE for PDFs;
instead of comparing true \revone{$X_{i}(z)$}{$p(z | x_{i})$} and estimated \revone{$x_{i}(z)$}{$p(z | x_{i}, \pi_{j})$} PDF values pointwise over $z$, the CDE loss evaluates expectations of the posterior PDFs of $z$ conditioned on photometric data \revone{$\xi$}{$x_{i}$} with respect to the \revone{}{population of} photometry \revone{$\Xi$}{$X = \{x_{i}\}$} and redshifts $Z$\revone{}{$= \{z_{i}\}$} of the test set data,
\begin{equation}
\mathcal{L}_{\mathrm{CDE}} \equiv \mathbb{E}_{X}\left[\int (p(z | x_{i}, \pi_{j}))^{2} dz\right] - 2\mathbb{E}_{X, Z}\left[p(Z | X)\right] + K ,
\label{eqn:cde}
\end{equation}
for constant $K$ dependent on the true PDFs (which is thus dropped in calculation since it is the same across estimates).
A more complete exposition of this metric can be found in \citet{Izbicki17}.

\end{description}

\end{document}